\documentclass[aps,pre,reprint,superscriptaddress,footinbib,showpacs,floatfix]{revtex4-1}

\usepackage{comment}
\usepackage{graphicx}
\usepackage{latexsym}
\usepackage{xcolor}
\usepackage{amsmath}
\usepackage[caption=false]{subfig}
\usepackage{multirow}
\usepackage{booktabs}
\usepackage{textcomp}
\usepackage{placeins} % Makes \FloatBarrier available!
\usepackage{float}
\usepackage{ulem} % Use \sout{} for strikethrough text.
\usepackage{amsfonts,amssymb}
\usepackage{bbold}
\usepackage{accents}
\graphicspath{{images/}}

\renewcommand{\vec}[1]{\mathbf{#1}}

\newcommand{\be}{\begin{equation}}
\newcommand{\ee}{\end{equation}}
\newcommand{\bea}{\begin{eqnarray}}
\newcommand{\eea}{\end{eqnarray}}

\begin{document}

\title{Length scale dependent elasticity in DNA from coarse-grained and
all-atom models}

\author{Enrico Skoruppa}
\affiliation{Laboratory for Soft Matter and Biophysics, KU Leuven,
Celestijnenlaan 200D, 3001 Leuven, Belgium}
\author{Aderik Voorspoels}
\affiliation{Laboratory for Soft Matter and Biophysics, KU Leuven,
Celestijnenlaan 200D, 3001 Leuven, Belgium}
\author{Jocelyne Vreede}
\affiliation{Van 't Hoff Institute for Molecular Sciences,
University of Amsterdam, Science Park 904, 1098 XH Amsterdam}
\author{Enrico Carlon}
\affiliation{Laboratory for Soft Matter and Biophysics, KU Leuven,
Celestijnenlaan 200D, 3001 Leuven, Belgium}

\date{\today}

\begin{abstract}
The mechanical properties of DNA are typically described by elastic
theories with purely local couplings (on-site models).  We discuss and
analyze coarse-grained (oxDNA) and all-atom simulations, which indicate
that in DNA distal sites are coupled. Hence, off-site models provide
a more realistic description of the mechanics of the double helix.
We show that off-site interactions are responsible for a length scale
dependence of the elasticity, and we develop an analytical framework to
estimate bending and torsional persistence lengths in models including
these interactions.  Our simulations indicate that off-site couplings
are particularly strong for certain degrees of freedom, while they are
very weak for others.  If stiffness parameters obtained from DNA data
are used, the theory predicts large length scale dependent effects for
torsional fluctuations and a modest effect in bending fluctuations,
which is in agreement with experiments. 
\end{abstract}

\maketitle

\section{Introduction}

Mechanical properties of DNA strongly influence how the double helix
performs its various tasks in the cell, where it is often bent and twisted
\cite{agga20}. Computer simulations have been playing an increasingly
important role in understanding these properties.  Depending on the
length scale relevant to the particular issue at hand and the level of
detail required, simulations of either atomistic \cite{lank03, lank00,
lave10, noy12, pasi17, cler18, vela20} or coarse-grained resolution
\cite{samb09, dans10, ould10, sulc12, fred14, fosa16, skor17, chak18,
li18, skor18, henr18, cara19} can be employed. It is well documented that
at length scales beyond a couple of helical repeat lengths the mechanical
response of DNA is well described by continuous elastic models, such
as the Twistable Worm-like Chain (TWLC) \cite{nels08}.  At these length
scales sequence effects are averaged out and DNA can be described as
a homogeneous chain composed of a sequence of elastic elements coupled
via strictly nearest-neighbor interactions. We will refer to this type
of models as on-site models.  Contrarily, at shorter distances this
simple approach breaks down as sequence specificity starts to dominate
the elastic behavior and the assumption of coupling locality does no
longer hold.  The former issue is well-documented - several studies have
shown that DNA elasticity at the base pair level is strongly dependent
on the involved type of nucleotides \cite{lank00,lave10,noy12} - while
the latter issue is the main concern of this paper. Couplings beyond
nearest-neighbors have been observed in all-atom simulations \cite{lank09}
as well as in coarse-grained models \cite{skor17}, suggesting that on-site
models provide an approximate description of DNA elasticity. However,
these effects are typically not accounted for in models of DNA mechanics.
In this work we investigate these non-local interactions
and explore their connection to length scale dependent elasticity.

We present here the results of simulations conducted with a homogeneous
coarse-grained DNA model and an all-atom model for which we average
over different sequences. The central quantity in our analysis is
the set of momentum space stiffness matrices, that capture the linear
response of the model at all length scales and present a convenient way
to quantify the effect of beyond nearest neighbor interactions. Here,
we do not discuss extreme bendability at short scales and kinking, which
would require an energetic model including beyond-harmonic interactions
(for a recent study of kinking, see e.g.\ Ref.~\cite{schi18}).

Although our focus here is DNA, it turns out that length
scale dependent elasticity can also be understood in simpler
systems. Therefore, we start our discussion introducing a ``toy" model
(Section~\ref{sec:toy}). This model shows a length scale dependent elastic
stiffness (Eq.~\eqref{toy:Km_int}) and the exponential decay of a local
perturbation (Eq.~\eqref{toy:allostery}) which are also found in DNA. The
advantage is that the toy model is simpler and perhaps more intuitive
to understand. In addition, several quantities can be computed exactly.
In Section~\ref{sec:DNA} the formalism introduced for the simple model is
transferred to our three dimensional model for DNA.  Numerical results
obtained with the coarse grained and atomistic model are presented
in Section~\ref{sec:oxDNA}.  Finally, in Section~\ref{sec:conclusion}
we discuss the results obtained and link our findings to experimental
observations.

%%%%%%%%%%%%%%%%%%%%%%%%%%%%%%%%%%%%%%%%%%%%%%%%%%%%%%%%%%%%%%%%%%%%  
\begin{figure}[b]
\includegraphics[width=0.45\textwidth,angle=0]{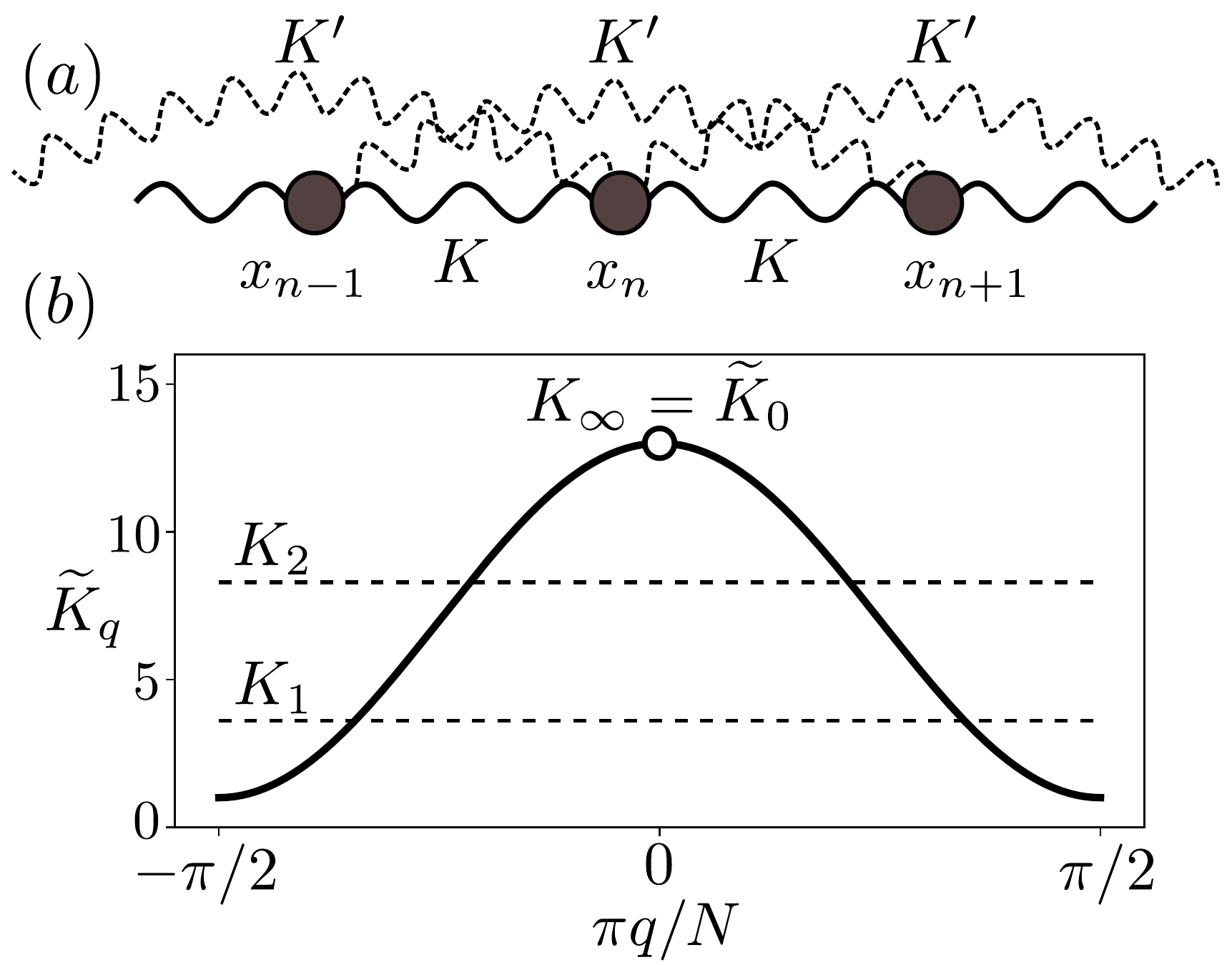}
\caption{(a) ``Toy" model of length scale dependent elasticity
consisting of a linear chain with neighbors and next-neighbors springs
with stiffnesses $K$ and $K'$, respectively (Eq.~\eqref{toy:mod1}).
(b) Momentum space stiffness of the model \eqref{toy:Ktilde} for
$K=1$ and $K'=3$. The one-step $K_1$, two-step $K_2$ and asymptotic
stiffnesses $K_\infty = \widetilde{K}_0$ (Eqs.~\eqref{K1}, \eqref{K2}
and \eqref{Kexp}) are shown.  In the case shown here ($K'>0$) the system
is softer at short scales: $K_1 < K_2 < \ldots < K_\infty$.}
\label{fig:ladder}
\end{figure}
%%%%%%%%%%%%%%%%%%%%%%%%%%%%%%%%%%%%%%%%%%%%%%%%%%%%%%%%%%%%%%%%%%%%  

\section{Linear elastic chain with next nearest-neighbor coupling}
\label{sec:toy}

In order to illustrate the effect of beyond-nearest-neighbor couplings
and the procedure of analyzing length-dependent elasticity we first consider
a one dimensional ``toy" model of a linear elastic chain with next
neighbors couplings.

%The toy model is shown in Fig.~\ref{fig:ladder}(a)) and consists
%of an elastic chain of masses
%%  To elaborate on this further we work out the elasticity of a linear
%with two types of springs with stiffnesses $K$ and $K'$ and rest
%lengths $a$ and $2a$, acting respectively between nearest-neighbors
%and next-nearest neighbors. 
%The energy of the system, in units of $k_BT$,
%is given by:
%\begin{equation}
%\beta E = \frac{K}{2} \sum_{n=0}^{N-1} (x_{n+1} - x_{n} - a)^2 + 
%\frac{K'}{2} \sum_{n=0}^{N-1} (x_{n+2} - x_{n} - 2a)^2 
%\label{toy:mod1}
%\end{equation}
%with $\beta = 1/k_BT$, $x_n$ the positions of the masses and
%where we have used the following boundary conditions $x_{N+1} = x_0 +
%(N+1)a$. 
 
This model (illustrated in Fig.~\ref{fig:ladder}(a)) consists 
of an elastic chain of $N$ masses located at positions $x_n$, which are 
subjected to periodic boundary conditions ($x_{N+1} = x_0 +
(N+1)a$). These boundary conditions are formally necessary for our formalism,
however their violation merely constitutes a finite size effect
that will vanish for sufficiently large $N$. 
Interactions between the masses are mediated by two types of 
springs with stiffnesses $K$ and $K'$ and rest
lengths $a$ and $2a$, acting respectively between nearest-neighbors
and next-nearest neighbors. Accordingly, the energy of the system - 
in units of $k_BT$ - is given by 
\begin{equation}
\beta E = \frac{K}{2} \sum_{n=0}^{N-1} (x_{n+1} - x_{n} - a)^2 + 
\frac{K'}{2} \sum_{n=0}^{N-1} (x_{n+2} - x_{n} - 2a)^2,
\label{toy:mod1}
\end{equation}
with $\beta = 1/k_BT$. The minimal energy configuration of the system
is $x_n = x_0 + na$. We are interested in the stretching fluctuations
at different length scales, as captured by the m-step fluctuations
\begin{equation}
\langle (x_{m} - x_0 - ma)^2 \rangle = \frac{m}{K_m},
\label{toy:defKm}
\end{equation}
for which we define an effective spring constant $K_m$.  In absence of
next-nearest neighbor couplings ($K'=0$) one simply finds $K_m=K$, as
the mean-squared extension of $m$ independent springs is just $m$ times
the extension of a single spring, which yields the stated relation by
virtue of the equipartition theorem. As we shall show, in the case $K'
\neq 0$ the spring constant $K_m$ depends on $m$, indicating a length
dependent elasticity.

For the calculation of $K_m$ we define the displacement from the springs
rest length as $u_n \equiv x_{n+1} - x_{n} - a$, such that \eqref{toy:mod1}
becomes
\begin{eqnarray}
\beta E &=& \frac{K}{2} \sum_{n=0}^{N-1} u_n^2 + 
\frac{K'}{2} \sum_{n=0}^{N-1} (u_{n+1} + u_n)^2.
\label{eq:discrete}
\end{eqnarray}
We introduce the discrete Fourier transform of the displacements
\begin{equation}
{\cal U}_q = \sum_{n=0}^{N-1} e^{-2\pi i qn/N} \, u_n,
\label{def:uq}
\end{equation}
%  with $q=-N/2,-(N-1)/2,\ldots N/2-1$ (assuming $N$ even) 
%  \es{[Perhaps I'm missing something, but isn't it more convenient to
%  have an even number of non-zero q modes?]}. 
with $q=-(N-1)/2,-(N-3)/2,\ldots (N-1)/2$ (assuming $N$ odd) 
referred to as momentum here. 
%The inverse formula is
Accordingly, the inverse Fourier transform is given by
\begin{equation}
{u}_n = \frac{1}{N} %\sum_{q=-(N-1)/2}^{(N-1)/2} 
\sum_q
e^{2\pi i qn/N} \, {\cal U}_q,
\label{def:un}
\end{equation}
where the sum runs over the above given values of $q$.
Since the $u_n$ are real variables we have ${\cal U}_q^* = {\cal U}_{-q}$.
In momentum space the energy then becomes
\begin{equation}
\beta E = \frac{1}{2N} \sum_q \widetilde{K}_q |{\cal U}_q|^2.
\label{toy:Eq}
\end{equation}
%  For simplicity we have omitted the range of $q$-values in the sum.
%  \es{[Do you think this needs to be mentioned? That seems totally standard
%  to me]}
The stiffness of the mode with momentum $q$ obeys 
\begin{equation}
\widetilde{K}_q \equiv  K+ 4 K' \cos^2 \frac{\pi q}{N}.
\label{toy:Ktilde}
\end{equation}
From here one can easily deduce the stability condition of the system:
$\widetilde{K}_q > 0$ for all $q$ requires $K>0$ and $K' > -K/4$.
Figure~\ref{fig:ladder}(b) shows $\widetilde{K}_q$ for $K=1$ and $K'=3$.

The equipartition theorem, applied to \eqref{toy:Eq} gives
\begin{equation}
\langle {\cal U}_q \, {\cal U}_{q'} \rangle = 
N\widetilde{K}_q^{-1} \, \delta_{q,-q'},
\label{toy:ave_Uq2}
\end{equation}
where $\delta_{n,k}$ is the Kronecker delta. 
{Moreover, collective m-step fluctuations can be expressed as
\begin{equation}
x_m -x_0 -ma = \sum_{n=0}^{m-1} u_n = \frac{1}{N} 
\sum_q %f_m(q) \, 
\frac{\sin \frac{\pi qm}{N}}{\sin \frac{\pi q}{N}}
e^{i\pi q(m-1)/N} {\cal U}_q.
\label{toy:summ}
\end{equation}
%  with
%  \begin{equation}
%  f_m(q) =\frac{\sin \frac{\pi qm}{N}}{\sin \frac{\pi q}{N}}.
%  \label{toy:fmq}
%  \end{equation}
%  \textcolor{blue}{\textit{[It seems to me that we are not using this function further down the text, hence it 
%  seems redundant to even define it. Alternatively, we could replace all occurances of the factor
%  replace with $f_m$.]}}
%  

%  \sout{The sum in \eqref{toy:summ} is extended to the whole momentum space
%  and it is modulated by the function $f_m(q)$.} \es{(redundant)} This is an oscillatory
%  decaying function from its maximum value $f_m(q=0)=m$. 
%  
Combining
\eqref{toy:defKm}, \eqref{toy:ave_Uq2} and \eqref{toy:summ} we find
\begin{equation}
\frac{m}{K_m} = 
%  \frac{1}{N^2} \sum_q f^2_m(q) \langle |{\cal U}_q|^2 \rangle =
\frac{1}{N^2} \sum_q 
\frac{\sin^2 \frac{\pi qm}{N}}{\sin^2 \frac{\pi q}{N}}
\langle |{\cal U}_q|^2 \rangle =
%
%  \frac{1}{N} \sum_q \frac{f^2_m(q)}{\widetilde{K}_q}.
\frac{1}{N} \sum_q \frac{\sin^2 \frac{\pi qm}{N}}
{ \widetilde{K}_q \sin^2 \frac{\pi q}{N}}.
\label{toy:Km}
\end{equation}
In the limit $N \to \infty$ one can replace the discrete sum 
with an integral
\begin{eqnarray}
\frac{1}{K_m} &=& \frac{1}{m \pi} \int_{-\pi/2}^{\pi/2} 
\frac{\sin^2 my}{\sin^2 y} \frac{dy}{K+4K'\cos^2 y},
\label{toy:Km_int}
\end{eqnarray}
where we defined $y \equiv \pi q/N$ and used \eqref{toy:Ktilde}.
For $m=1$ and $m=2$ a straightforward calculation shows that
\begin{eqnarray}
&&K_1 = \sqrt{K(K+4K')} 
\label{K1}\\
&&K_2 = \frac{2K'\sqrt{K+4K'}}{\sqrt{K+4K'}-\sqrt{K}}.
\label{K2}
\end{eqnarray}
In the asymptotic limit of large $m$ the factor $\sin^2(my)/ \sin^2 y$
in \eqref{toy:Km_int} becomes increasingly peaked around $y=0$.
Expanding $1/(K+4K' \cos^2 y)$ to lowest orders in $y$ we obtain
in the case $m \gg 1$
\begin{eqnarray}
&&K_m = \widetilde{K}_0 
- \frac{4 K' \log2}{m} + {\cal O}\left(\frac{1}{m^2} \right),
\label{Kexp}
\end{eqnarray}
where we used
\begin{eqnarray}
\int_{-\pi/2}^{\pi/2} \frac{\sin^2 my}{\sin^2 y} \, dy
&=& {m \pi},
\label{toy:int_sinmsin}
\end{eqnarray}
and 
\begin{eqnarray}
\int_{-\pi/2}^{\pi/2} \frac{y^2 dy}{\sin^2 y}
= \pi\log4. 
\end{eqnarray}
Equations \eqref{K1}, \eqref{K2} and \eqref{Kexp} show that the stiffness
of the chain depends on the length scale at which fluctuations are
observed. In the case $K'>0$ one finds $K_1 < K_2 < \ldots < K_\infty$,
eg.\ the chain becomes increasingly stiffer at longer length scales
(Fig.~\ref{fig:ladder}(b)). The behavior is the opposite if $K' <0$:
the chain is softer at longer distances $K_1 > K_2 > \ldots > K_\infty$.
As $m$ increases the contribution of large momenta to $K_m$ gradually
diminishes, until finally only the zero-momentum component ($q=0$)
contributes to the asymptotic stiffness $K_\infty=\widetilde{K}_0$. In the
opposite limit ($m=1$) $K_1$ becomes the harmonic mean of the momentum
domain stiffnesses $\widetilde{K}_q$.  Recall that the harmonic mean of
$N$ numbers $\omega_i$ with $i=1$, $2$ \ldots $N$ is defined as
\begin{eqnarray}
\langle \omega \rangle_h = 
\left( \frac{1}{N} \sum_i \frac{1}{\omega_i} \right)^{-1}.
\label{toy:HM}
\end{eqnarray}
We consider now the effect of a local perturbation stretching one of the
springs (say $u_0$). This can be achieved by imposing a local force $f >0$ 
on the selected degree of freedom such that the energy becomes
\begin{equation}
\beta E_f = \beta E - \beta f u_0
= \frac{1}{2N} \sum_q \widetilde{K}_q |{\cal U}_q|^2 - 
\frac{\beta f}{N} \sum_q {\cal U}_q,
\label{toy:localf}
\end{equation}
with $\beta E$ the unperturbed energy \eqref{toy:Eq}. The force
stretches all modes to a non-zero average 
\begin{eqnarray}
\langle {\cal U}_q \rangle = \frac{\beta f}{\widetilde{K}_q}.
\label{toy:shift_Uq}
\end{eqnarray}
The inverse Fourier transform then gives (for details see
Appendix~\ref{app:allostery})
\begin{eqnarray}
\langle u_m \rangle &=& 
\frac{\beta f}{N}  \sum_q \frac{e^{2iqm/N}}{\widetilde{K}_q}
= \frac{\beta f}{\pi} \int_{-\pi/2}^{\pi/2} \frac{e^{2iym} \, dy}{K+4K'
\cos^2 y} \nonumber\\
&=& \frac{\beta f}{K_1} \left[ - \text{sgn} (K') \right]^m
\, e^{-m/l_\text{A}},
\label{toy:allostery}
\end{eqnarray}
with $m>0$, sgn denoting the signum function and
\begin{equation}
\frac{1}{l_\text{A}} = -\log 
\frac{|K_2-K_1|}{K_2}.
\label{toy:def_xi}
\end{equation}
Here $K_1$ and $K_2$ are the one-step and two-step stiffnesses
defined in \eqref{K1}, \eqref{K2}. We note that for $m=0$ we get from
\eqref{toy:allostery} $K_1 \langle u_0 \rangle = \beta f$, showing again
that $K_1$ is the stretching stiffness between neighboring sites. If
$K'>0$, the quantity $\langle u_m \rangle$ has an oscillatory decay,
which can be easily understood from the coupling term $K' u_n u_{n+1}$,
that contributes negatively if neighboring $u_n$ have opposite signs.
The same reasoning explains the monotonic decay if $K' < 0$.  Note that
in absence of length scale dependence, which means that $K_m$ does not
depend on $m$, one has $l_\text{A}=0$.  Hence, in that case, a local
perturbation does not affect flanking springs.

%%%%%%%%%%%%%%%%%%%%%%%%%%%%%%%%%%%%%%%%%%%%%%%%%%%%%%%%%%%%%%%%%%%%  
\begin{figure}[t]
\includegraphics[width=0.48\textwidth,angle=0]{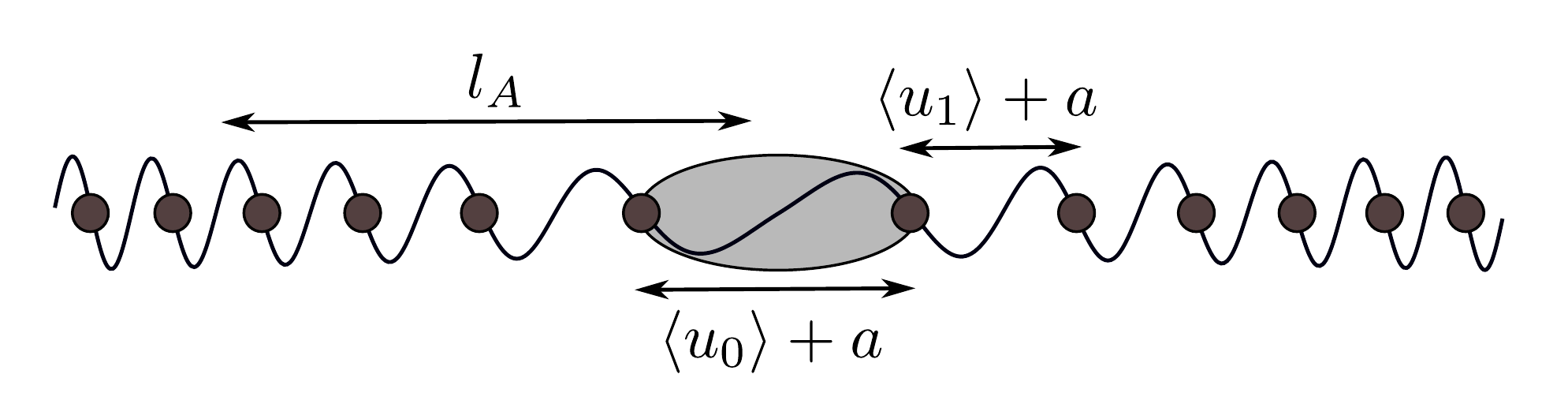}
\caption{Schematic illustration of the effect of a local perturbation
at site $n=0$ resulting in an exponentially decaying stretching profile
$\langle u_m \rangle$, see Eq.~\eqref{toy:allostery}.  This depiction
represents the case $K'<0$, where the stretching decays monotonically
(for the sake of clarity we do not show next-neighbors springs).}
\label{fig:allostery}
\end{figure}
%%%%%%%%%%%%%%%%%%%%%%%%%%%%%%%%%%%%%%%%%%%%%%%%%%%%%%%%%%%%%%%%%%%%  

To conclude the analysis of the model we remark that while our discussion
here was limited to interactions ranging to next-nearest neighbors,
i.e. involving just two spring constants ($K$and $K'$), the same formalism
is directly applicable to systems involving further ranging interactions.
In that case \eqref{toy:Km} and \eqref{toy:allostery} remain valid,
but $\widetilde{K}_q$ will assume a more complicated form.

\section{DNA elasticity in momentum space}
\label{sec:DNA}

In our coarse-grained description of DNA any configuration of a molecule
consisting of $N+1$ base pairs is fully described by a set of $N+1$
orthonormal triads $\widehat{\cal T}_n = ( \widehat{\bf f}_n \widehat{\bf
v}_n \widehat{\bf u}_n )$, where $\widehat{\bf f}_n,\widehat{\bf v}_n$
and $\widehat{\bf u}_n$ are unit vectors capturing the local geometry of
the base pair. We define $\widehat{\bf u}_n$ to be the local tangent and
$\widehat{\bf v}_n$ to connect the two oppositely running backbones such
that the remaining vector $\widehat{\bf f}_n = \widehat{\bf v}_n \times
\widehat{\bf u}_n$ points towards the major groove (in the literature
this frame is indicated also as $(\widehat{\bf e}_1 \widehat{\bf e}_2
\widehat{\bf e}_3 )$ \cite{mark94,skor17}, here we use a different
notation to avoid double indexing $\widehat{\bf e}_{1,n}$).  The spacial
configuration of the molecule is given by the set of points connected
by the vectors $a \widehat{\bf u}_n$, where $a$ is the distance between
consecutive base pairs. We assume this distance to be the constant
value $a=0.34$ nm. For simplicity this description ignores stretching
deformations. However, such could easily be included by replacing the
connection vector $a \widehat{\bf u}_n$ by a variable $3$-component
vector.

Up to a global rotation a particular chain configuration is fully captured
by the set of rotations that map each triad onto its consecutive triad, as
illustrated in Fig.~\ref{fig:dna_frames}. It is convenient to parametrize
these rotations by the corresponding Euler vectors $\vec\Theta$,
i.e. the vectors parallel to the rotation axis with magnitude $\Theta =
|\vec\Theta|$ equal to the rotation angle. In order to link the vector
components to the local geometry we express it in the basis of the local
material frame
\begin{equation}
\vec\Theta_n = 
a \tau_n \widehat{\bf f}_n + a \rho_n \widehat{\bf v}_n + 
a (\Omega_n+\omega_0) \widehat{\bf u}_n.
\label{DNA:defTheta}
\end{equation} 
The components $\tau$ and $\rho$ denote the two bending modes commonly
referred to as tilt and roll \cite{lave09}, quantifying local
bending over the axes $\widehat{\bf f}_n$ and $\widehat{\bf v}_n$
respectively.  The total twist $\Omega_n+\omega_0$ (rotation around
$\widehat{\bf u}_n$) has two components: $\Omega_n$ is the excess
twist and $\omega_0 = 1.75\,\text{nm}^{-1}$ the intrinsic twist of
the double helix, corresponding to one turn of the helix every $10.5$
base pairs. The deformation densities $\tau_n$, $\rho_n$ and $\Omega_n$
of \eqref{DNA:defTheta} have the dimension of inverse lengths and are
expressed in nm$^{-1}$, while $a\tau_n$, $a\rho_n$ and $a\Omega_n$
are dimensionless and express rotation angles in radians.

The configuration $\tau_n=\rho_n=\Omega_n=0$ (all $n$) corresponds to a
straight twisted rod with intrinsic twist $\omega_0$, which is assumed
to be the ground state of the system. Any deformation away from this
state will be associated with a certain free energy. Expanding this
free energy to lowest non-vanishing order around the ground state then
corresponds to a regime of linear elasticity. In this work we limit our
discussion to this regime. It is customary to describe DNA elasticity
using on-site models, e.g. without interactions between neighboring
sites. For instance, the Marko-Siggia model \cite{mark94} is defined as
\begin{equation}
\beta E = \frac{a}{2} \sum_n \left( A^t \tau_n^2 + A^r \rho_n^2+
C  \Omega_n^2 + 2 G \rho_n \Omega_n \right),
\label{dna:local}
\end{equation}
where $A^t$, $A^r$, $C$ and $G$ are stiffness parameters (we
neglect in this description sequence dependent effects and use
constant stiffnesses). Besides the individual stiffnesses of tilt
($A^t$), roll ($A^r$) and twist ($C$), the model \eqref{dna:local}
is characterized by a non-vanishing twist-roll coupling ($G$), as
expected from the symmetry of the molecule \cite{mark94}. The effects
of this coupling in the conformations of a DNA molecule were discussed
recently in \cite{skor18,cara19,nomi19a}. 

%%%%%%%%%%%%%%%%%%%%%%%%%%%%%%%%%%%%%%%%%%%%%%%%%%%%%%%%%%%%%%%%%%%%  
\begin{figure}[t]
\includegraphics[width=0.48\textwidth]{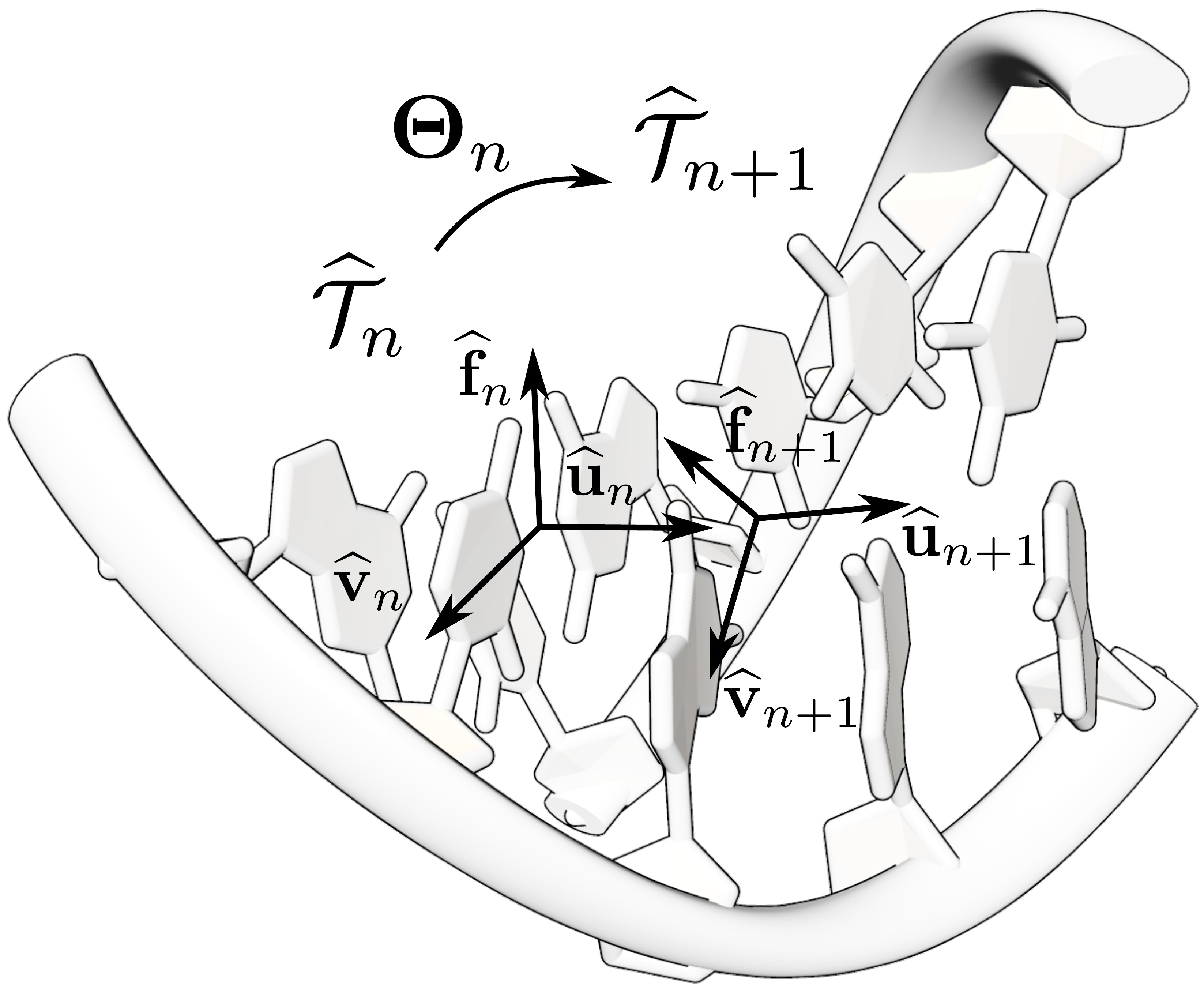}
\caption{
Mapping of a DNA configuration into a rigid basepair
representation~\cite{lank09} that consists of a series of triads
each attached to a single basepair, capturing the local geometry of
the molecule. These triads are constructed from a set of 3 mutually
orthogonal unit vectors $\widehat{\cal T}_n = ( \widehat{\bf f}_n
\widehat{\bf v}_n \widehat{\bf u}_n )$, where $\widehat{\bf u}_n$ is
the local tangent, $\widehat{\bf v}_n$ connects the two backbones and
$\widehat{\bf f}_n$ points towards the major groove. Deformation of the
chain are parametrized by the rotation vectors $\vec\Theta_n$ rotating
the triads $\widehat{\cal T}_n$ into their sequentially adjacent triads
$\widehat{\cal T}_{n+1}$.}
\label{fig:dna_frames}
\end{figure}
%%%%%%%%%%%%%%%%%%%%%%%%%%%%%%%%%%%%%%%%%%%%%%%%%%%%%%%%%%%%%%%%%%%%  

We generalize the elastic model to allow for interactions between
further neighbors employing a matrix representation 
\begin{equation}
\beta E = \frac{a}{2} \sum_n \sum_m
\vec{\Delta}_n^\intercal M_m \vec{\Delta}_{n+m},
\label{dna:non_local}
\end{equation}
with $\vec{\Delta}_n^\intercal = (\tau_n, \rho_n, \Omega_n)$ and where
the $M_m$ are $3 \times 3$ matrices describing the couplings between sites
separated by $m$ steps. Stability of the model requires the on-site
matrices $M_0$ to be positive definite. For homogeneous directionally
invariant chains the general form of the matrices $M_m$ can be deduced
from symmetry considerations.  Reversal of the curvilinear coordinate
system, i.e. a definition of the $\vec\Theta_n$ in backwards-sense rather
than forwards-sense results in the same stiffness matrices, however for
a given configuration this sense-reversal transformation leads to the
transformation $\vec{\Delta}_n^\intercal = (\tau_n, \rho_n, \Omega_n)
\to (-\tau_n, \rho_n, \Omega_n) = \vec{\bar{\Delta}}_n^\intercal$
\cite{mark94}. Since this coordinate transformation cannot change the
energy we see that for every $m$
\begin{equation}
\vec{\Delta}_n^\intercal M_m \vec{\Delta}_{n+m} = 
\vec{\bar{\Delta}}_{n+m}^\intercal M_m \vec{\bar{\Delta}}_{n}.
\label{dna:symmetry}
\end{equation}
This implies that all off-diagonal terms in $M_m$ involving $\tau$, have
to be anti-symmetric, while the remaining coupling (between $\rho$ and
$\Omega$) is required to be symmetric.  Hence, for homogeneous chains,
the most general form of the matrices $M_m$ is
\begin{equation}
M_m =
\begin{pmatrix}
\phantom{-}A_m^t & A_m^{tr} & B_m\\
-A_m^{tr} & A_m^r & G_m\\
-B_m & G_m   & C_m\\
\end{pmatrix}.
\label{dna:Mm_hom}
\end{equation}
For example, the coupling $A_m^{tr}$ gives rise to terms of the form 
\begin{equation}
\frac{1}{2} \sum_{n} {A}^{tr}_{m} 
\left( {\tau}_{n} {\rho}_{n+m} - {\rho}_{n} {\tau}_{n+m} \right).
\end{equation}

This symmetry consideration implies that for homogeneous on-site models,
i.e. $M_m =0$ for $m \geq 1$, the most general form of the free energy
density (in the regime of linear elasticity) is given by the afore
mentioned Marko-Siggia model~\eqref{dna:local}. In matrix representation 
this corresponds to a $M_0$ of the form \eqref{dna:Mm_hom} with $A^{tr}_0=B_0=0$.

% I HAVE NOTICED WE HAVE TOO MANY FOOTNOTES....
%   which has the matrix representation 
%  \footnote{Note that positive definite matrices
%  are symmetric.
%  \begin{equation}
%  M_0 =
%  \begin{pmatrix}
%  A^t & 0 & 0\\
%  0 & A^r & G\\
%  0 & G   & C\\
%  \end{pmatrix}.
%  \label{dna:MS}
%  \end{equation}
%  }

%  \esr{Perhaps then we should also remove the 
%  ``which has the matrix representation''. Otherwise the footnote becomes 
%  part of the sentence. }

%  I FEEL THIS BREAKS THE LINE OF REASONING HERE
%  The linear elastic chain of Sec.~\ref{sec:toy} can also be cast
%  in the form \eqref{dna:non_local} \es{assuming the matrices $M_m$
%  to be one-dimensional}. Developing the square $(u_n+u_{n+1})^2$ in
%  \eqref{eq:discrete} one finds $M_0 = K + 2K'$, $M_1=2K'$ and $M_m =0 $
%  for $m \geq 2$.

We can rewrite the model \eqref{dna:non_local} in momentum space as
\begin{equation}
\beta E = \frac{a}{2N} \sum_q \widetilde{\vec{\Delta}}_q^\dagger
\widetilde{M}_q \widetilde{\vec{\Delta}}_{q},
\label{dna:ham_disc}
\end{equation} 
where $\widetilde{\vec{\Delta}}_q$ and $\widetilde{M}_q$ are the
Fourier transform of $\vec{\Delta}_n$ and $M_m$, respectively,
and \textsuperscript{$\dagger$} indicates the conjugate transpose.
Stability of the model requires each of the Hermitian
\footnote{
$\tilde{M}_q$ is Hermitian because
$\tilde{\vec{\Delta}}_q \tilde{M}_q
\tilde{\vec{\Delta}}_{q}$ is real for every $q$.}
matrices $\widetilde{M}_q$ to be be positive definite, i.e. that all
eigenvalues are positive. As indicated in \eqref{dna:Mm_hom} the
matrices $M_m$ may contain symmetric and anti-symmetric components.
%  By fouriertransforming the symmetric and anti-symmetric matrices separately
%  one can directly deduce that they give rise to real and imaginary entries
%  in $\widetilde{M}_q$ respectively.  Hence for homogeneous chains the
%  most general form of the momentum space coupling matrices is
% PERHAPS SIMPLE FORMULATION
Fourier transformation in $m$ of the matrices \eqref{dna:Mm_hom} gives
\begin{equation}
\widetilde{M}_q =
\begin{pmatrix}
\phantom{-}\widetilde{A}^t_q & i \widetilde{A}^{tr}_q & i\widetilde{B}_q\\
-i \widetilde{A}^{tr}_q & \widetilde{A}^r_q & \widetilde{G}_q\\
-i\widetilde{B}_q & \widetilde{G}_q   & \widetilde{C}_q\\
\end{pmatrix},
\label{DNA:stiff}
\end{equation}
where all entries $\widetilde{A}^t_q$, $\widetilde{A}^r_q$,
$\widetilde{A}^{tr}_q$, $\widetilde{B}_q$, $\widetilde{C}_q$,
and $\widetilde{G}_q$ are real variables. The off-diagonal terms
$\widetilde{A}^{tr}_q$, $\widetilde{B}_q$ are odd functions of $q$
(e.g. $\widetilde{A}^{tr}_{-q} = -\widetilde{A}^{tr}_q$), while all
other terms are even functions of $q$.

%\esr{I suggest we discuss this and then I absorb the final lines of this 
%subsection into this general discussion and construct the text such 
%that in \eqref{DNA:stiff} we indeed observe this functional form.} 

The advantage of the momentum space representation is that modes with
different $q$ are independent (except for the coupling between $q$
and $-q$).  Strictly speaking, this is valid only if periodic boundary
conditions are imposed such that full translational invariance is
achieved.  In absence of that, some boundary terms will appear, which,
however, will be negligible for sufficiently large $N$.

Given an ensemble of deformation vectors $\vec{\Delta}_n$ the 
stiffness matrices can be obtained from the relation \cite{olso98}
\begin{eqnarray} 
\langle \widetilde{\vec{\Delta}}_{q} 
\widetilde{\vec{\Delta}}_{q}^\dagger \rangle &=& \frac{N} {a}
\widetilde{M}_q^{-1} ,
\label{DNA:cov} 
\end{eqnarray}
where the $3 \times 3$ covariance matrix $\langle
\widetilde{\vec{\Delta}}_{q} \widetilde{\vec{\Delta}}_{q}^\dagger \rangle$
is constructed from the ensemble averages of the products of the three
components of the vector $\widetilde{\vec{\Delta}}_q^\intercal =
(\widetilde{\tau}_q, \widetilde{\rho}_q, \widetilde{\Omega}_q)$.
In the remainder of this section we discuss the consequences of
this model extension on various DNA properties: length dependence of
persistence lengths and decays of local perturbations.

%  , which may give rise to allostery.

\subsection{Twist persistence length}

The twist-correlation function is defined as
\begin{equation}
{\cal C}_{\mathrm{T}} (m) = \left\langle \cos \left( a \sum_{n=0}^{m-1}
\Omega_n \right) \right\rangle = \text{Re}
\left\langle e^{\,\, 
\displaystyle i a \sum_{n=0}^{m-1} \Omega_n}  \right\rangle,
\label{tw:correl}
\end{equation}
where $\text{Re}$ denotes the real part. We are interested in the
twist persistence length, which is the characteristic decay-length
of twist-correlations
\begin{equation}
\frac{1}{l_\text{T}} = -\frac{1}{ma}\, \log {\cal C}_{\mathrm{T}} (m).
\end{equation}
At this point, we present only a sketch of the calculation,
as it is totally analogous to that of the elastic chain example
discussed in detail in Sec.~\ref{sec:toy}. In like manner, we
rewrite the sum in \eqref{tw:correl} in momentum space using the
expression \eqref{toy:summ}. The variables $\widetilde{\Omega}_q$
for different momenta are independent hence the total average
\eqref{tw:correl} factorizes in terms of the form $\langle \exp (i
\alpha_q \widetilde{\Omega}_q + i \alpha_{-q}  \widetilde{\Omega}_{-q})
\rangle$ (it is convenient to group terms $q$ and $-q$ together). Using
the property of Gaussian variables
\begin{equation}
\left\langle e^{\displaystyle \pm i \alpha X}\right\rangle = 
e^{\displaystyle -\frac{\alpha^2}{2} \langle X^2 \rangle},
\end{equation}
we obtain in the limit $N \to \infty$
\begin{eqnarray}
\frac{1}{l_\text{T}}
&=& \frac{a}{2\pi m} \int_{-\pi/2}^{\pi/2} \frac{\sin^2 my}{\sin^2 y}
\frac{\langle |\widetilde{\Omega}_{q} |^2\rangle}{N} \, dy,
\label{DNA:lT}
\end{eqnarray}
which is analogous to \eqref{toy:Km_int} and where we again used $y \equiv
\pi q/N$. Just as in the example of Sec.~\ref{sec:toy} the integral is
dominated by smaller and smaller $y$ contributions as $m$ increases.
The asymptotic twist persistence length ($m \to \infty$) is finally
entirely governed by the zero-momentum component
\begin{eqnarray}
\frac{1}{l_\text{T}} =
\frac{a}{2N} \langle\widetilde{\Omega}_{0}^2\rangle.
\label{DNA:lTq0}
\end{eqnarray}

\subsection{Bending persistence length}

From the tangent-tangent correlation function
\begin{equation}
{\cal C}_{\mathrm{B}}(m) = 
%  \langle \widehat{\bf u}_0^\intercal \widehat{\bf u}_m \rangle =
%  \widehat{\bf u}_0^\intercal \left\langle {\cal R}_{m-1} {\cal R}_{m-2}
%  \ldots {\cal R}_1 {\cal R}_0 \right\rangle \widehat{\bf u}_0
\langle \widehat{\bf u}_0 \cdot \widehat{\bf u}_m \rangle
\label{DNA:uu}
\end{equation}
%  where we have used the relations $\widehat{\bf u}_{k+1} = {\cal R}_{k}
%  \widehat{\bf u}_{k}$ (see Fig.~\ref{fig:dna_frames}) to express
%  the correlation function as a product of rotations. 
%  The bending persistence length is given by
one obtains the bending persistence length
\begin{equation}
\frac{1}{l_\text{B}} = -\frac{1}{ma}\, \log {\cal C}_{\mathrm{B}} (m).
\label{DNA:deflB}
\end{equation}
The twist-correlation function could be expressed exactly in terms of
the deformation vectors $\vec{\Delta}_n$. However, establishing such a
connection for ${\cal C}_{\mathrm{B}}$ requires some approximations.
%  The main difficulty is the non-additivity of the bending densities
%  $\tau_n$ and $\rho_n$, i.e. the total bending between distant triads
%  cannot be simply calculated by summing up the respective components.
%  The predominant contribution of this non-additivity stems from the
%  intrinsic twist component $\omega_0$, which is typically significantly
%  larger than the components of $\vec{\Delta}_n$.
%  An estimate of mean square fluctuations of tilt, roll and twist
%  indeed indicates that $|\tau|, |\rho|, |\Omega| \lesssim \omega_0/10$.
%  \esr{If I calculate the square root of the variance of $\rho$ using $A_2
%  = 40$nm I arrive at $\omega_0 \approx 6\sqrt{ \langle \rho^2 \rangle}
%  = 6 \sqrt{1/(a A_2)} )$.  Perhaps change the 10 into a 5? }
%  
%  In Appendix~\ref{app:rot_matrices} we account for the non-additivity
%  arising from the intrinsic twist and find the expression
Under the assumption that the rotations connecting neighboring triads
are dominated by the intrinsic twist component $\omega_0$, we derived the
following expression for the bending persistence length (for details see
Appendix~\ref{app:rot_matrices})
\begin{equation}
\frac{1}{l_\text{B}} = \frac{a}{\pi m} 
\int_{-\pi/2}^{\pi/2} \frac{\sin^2 my}{\sin^2 y} \, 
\frac{\Psi_{q+\Delta q} + \Psi_{q-\Delta q}}{N} \, dy,
\label{DNA:lB}
\end{equation}
where we defined $\Delta q = N a \omega_0/(2\pi)$ and
\begin{equation}
\Psi_{q} \equiv
\frac{1-\cos (a \omega_0)}{2 (a \omega_0)^2}
\left\langle 
\left| \widetilde{\tau}_{q}\right| ^2 +
\left| \widetilde{\rho}_{q}\right| ^2 
\right\rangle.
\label{DNA:def_Psi}
\end{equation}
This relation resembles Eq.~\eqref{DNA:lT} with the difference that here
the $y(q)$ contributions of the momentum space bending deformations (tilt
and roll) are replaced by the mean of the shifted momenta $q \pm \Delta
q$. This stems from the fact that in order to appropriately connect local
bending deformations to the total deformation of a given multi-step
segment (say from $\widehat{\vec{u}}_0$ to $\widehat{\vec{u}}_m$)
one needs to rotate the local reference frames to unwind the intrinsic
helical twist.  $\Delta q$ is indeed the momentum shift associated with
the  DNA intrinsic twist.  As we integrate in the rescaled variable $y
= \pi q/N$, the momentum shift corresponds to $\Delta y = a \omega_0/2
\approx \pi/10.5$, e.g.\ approximately one tenth of the $y$ domain ($10.5$
is the number of base pairs for a full turn of the double helix). In the
limit $m \to \infty$ the $q=y=0$ term is selected from the integral, and
the asymptotic persistence length becomes (using \eqref{toy:int_sinmsin})
\begin{equation}
\frac{1}{l_\text{B}} =
\frac{1-\cos(a \omega_0)}{a \omega_0^2 N} 
\left\langle 
\left| \widetilde{\tau}_{\Delta q}\right| ^2 +
\left| \widetilde{\rho}_{\Delta q}\right| ^2
\right\rangle.
\label{DNA:lBq0}
\end{equation}

%  \subsection{Allostery}
\subsection{Local perturbations}

Repeating the procedure applied to the linear chain model of
Section~\ref{sec:toy} we add a local perturbation at a given site of the
DNA. This perturbation is introduced by means of generalized ``forces"
acting on the rotational degrees of freedom associated with that site -
again we choose the site $n=0$, but translational invariance implies
that the results are equally valid for any given site - so that the
energy becomes
\begin{eqnarray}
\beta E_{\vec{f}} &=& \beta E - 
%  \beta \left( f_\tau \tau_0 + f_\rho \rho_0 + f_\Omega \Omega_0 \right)
\beta \vec{f}^\intercal \vec\Delta_0
\nonumber \\
&=&
\frac{a}{2N} \sum_q 
\left( 
\widetilde{\vec \Delta}_q^\intercal - \frac{\beta}{a} \vec{f}^\intercal
\widetilde{M}_q^{-1}
\right)
\widetilde{M}_q
\left( 
\widetilde{\vec \Delta}_q - \widetilde{M}_q^{-1} \frac{\beta}{a} \vec{f}
\right)
\nonumber \\
&-& \frac{\beta^2}{2Na} \vec{f}^\intercal \widetilde{M}_q^{-1}  
%\widetilde{M}_q 
\vec{f}
\label{DNA:forces}
\end{eqnarray}
where $\beta E$ is the unperturbed energy \eqref{dna:ham_disc} and
$\vec\Delta_0^\intercal = (\tau_0,\rho_0,\Omega_0)$. The vector $\vec
f^\intercal = (f_\tau, f_\rho, f_\Omega)$ contains three components
coupling to tilt, roll and twist, respectively. These generalized
forces shift the average $\widetilde{\vec{\Delta}}_q$ to the
non-zero value
\begin{equation}
\langle \widetilde{\vec{\Delta}}_q^\intercal \rangle = 
\frac{\beta}{a} \vec{f}^\intercal \widetilde{M}_q^{-1},
\label{DNA:Deltaq}
\end{equation}
which is the equivalent of \eqref{toy:shift_Uq}. In the DNA case the
calculation involves the inversion of the $3 \times 3$ matrix
$\widetilde{M}_q$
\begin{equation}
\widetilde{M}_q^{-1} = \frac{\text{Adj} \left[\widetilde{M}_q \right]}
{\text{det}\, \widetilde{M}_q},
\label{DNA:invMq}
\end{equation}
where $\text{Adj}[.]$ denotes the adjoint matrix. Combining
\eqref{DNA:Deltaq} and \eqref{DNA:invMq} and performing the inverse
Fourier transform we obtain
\begin{eqnarray}
\langle \vec{\Delta}_m^\intercal \rangle &=& 
\frac{\beta}{\pi} \int_{-\pi/2}^{\pi/2} 
\frac{\vec{f}^\intercal \,\text{Adj}
 \left[\widetilde{M}_q \right]}
{\text{det}\, \widetilde{M}_q}
\, e^{2iym} \, dy,
\label{DNA:allostery}
\end{eqnarray}
which is analogous to Eq.~\eqref{toy:allostery}, derived for the toy
model. As in that case, Eq.~\eqref{DNA:allostery} gives rise to an
exponential decay for large $m$: $\langle \vec{\Delta}_m\rangle \sim
\exp(-ma/l_\text{A})$. The characteristic decay length $l_\text{A}$
is given by the poles closest to the real axis of the integrand (see
Appendix~\ref{app:allostery}).  We note that stability of the energy
\eqref{dna:ham_disc} requires $\text{det}\, \widetilde{M}_q > 0$ in the
real $q$ domain. Hence poles have necessarily an imaginary component
responsible for the exponential decay.  In practice this integral can
be evaluated numerically from empirically obtained $\widetilde{M}_q$.

\section{DNA elasticity in coarse-grained and all-atom models}
\label{sec:oxDNA}

We discuss and compare here the elasticity of the coarse grained DNA model
oxDNA \cite{ould10}, and of an all atom model. The main focus
is the calculation of $\widetilde{M}_q$ from which various quantities
are obtained, following the framework discussed in the previous Section.

\subsection{oxDNA}

The oxDNA model treats nucleotides as single rigid objects,
that mutually interact via multiple sites representing the
most significant inter-base interactions: backbone-connectivity,
base-pairing and base-stacking. These interactions are parametrized so
as to reproduce thermodynamical, structural and mechanical properties
of DNA \cite{ould10}. oxDNA has been used to study a broad range of
processes such as DNA-melting, -hybridization, -supercoiling, -looping,
DNA strand-displacement mechanisms , DNA gels, nanotubes and origami
\cite{srin13,schm13,mate15,roma15,enge18,desa20,chha20}. Here we focus
exclusively on oxDNA2 \cite{snod15}, a version of the model with
asymmetric major and minor grooves. We used the procedure outlined
in \cite{skor17} to map the oxDNA coordinates to orthonormal triads
$(\widehat{\vec f}_n  \widehat{\vec{v}}_n  \widehat{\vec{u}}_n )$
(Fig.~\ref{fig:dna_frames}). This mapping is not unique and a few
alternative definitions have been discussed in \cite{skor17}.  Differences
in triads are carried over the to rotational modes $\vec{\Delta}_n$, which
leads to slightly different elastic behavior.  However, we observe the
Fourier spectra of the couplings to exhibit the same general features.
In particular, alternative triads give the same behavior at small $q$
(same asymptotic elasticity) and follow the same trend from small to
large $q$ behavior. We will present here the results from triad2,
as defined in \cite{skor17}.

Using molecular dynamics trajectories of oxDNA2 (details about simulations
can be found in~\cite{skor17}) we computed the Fourier spectra of
the rotational deformations $\widetilde{\vec{\Delta}}_q^\intercal =
(\widetilde{\tau}_q, \widetilde{\rho}_q, \widetilde{\Omega}_q)$. The
stiffness matrices $\widetilde{M}_q$ were then obtained by utilizing
Eq.~\eqref{DNA:cov}. The matrix entries vs. rescaled momentum $y\equiv
\pi q/N$ are plotted in Fig.~\ref{fig:stiff_fourier}(a).  These matrices
indeed follow the structure \eqref{DNA:stiff} as predicted by the
symmetry consideration. The anti-symmetric components turn out to be
very small, with $\widetilde{B}_q$ virtually zero.  The only significant
off-diagonal term in oxDNA2 is the twist-roll coupling $\widetilde{G}_q$
\cite{skor17}. We note that $\widetilde{A}^r_q$, the roll stiffness
is very weakly dependent on $q$ as compared to the other entries. This
weak dependence indicates that the roll-roll interaction is dominated by
the on-site term $\rho_n^2$. The strong dependence on $q$ for tilt-tilt
and twist-twist terms implies significant contributions from off-site
interactions $\tau_n \tau_{n+m}$ and $\Omega_n \Omega_{n+m}$, with $m >0$.

%%%%%%%%%%%%%%%%%%%%%%%%%%%%%%%%%%%%%%%%%%%%%%%%%%%%%%%%%%%%%%%%%%%%  
\begin{figure}[t]
\includegraphics[width=0.52\textwidth]{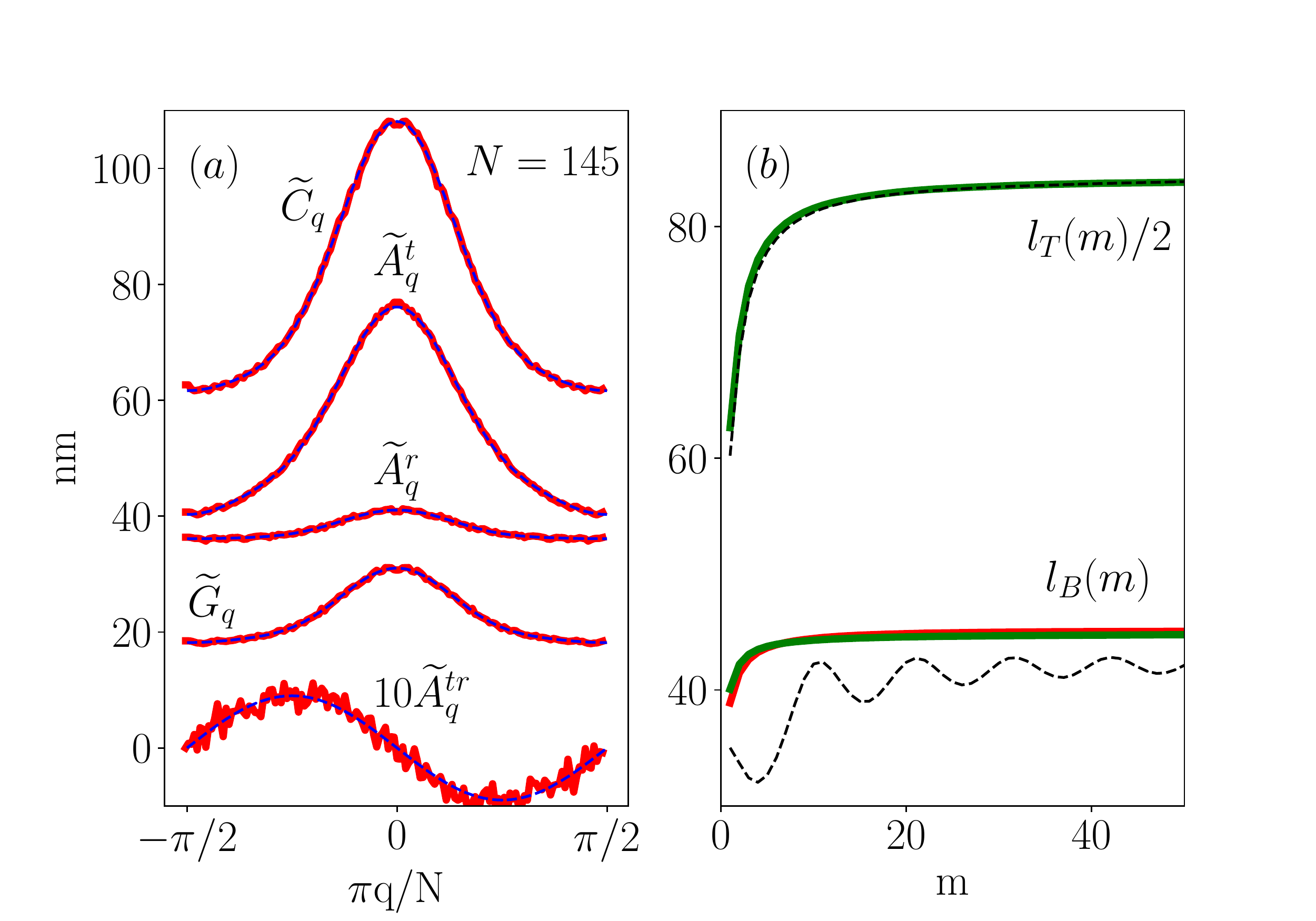}
\caption{(a) Red dots: Simulation data reporting the entries of the
stiffness matrix in momentum space $\widetilde{M}_q$ for oxDNA2 as
obtained from Eq.~\eqref{DNA:cov} for a sequence of length $150$.  In the
analysis two nucleotides at the two ends were eliminated, which gives
$146$ triads and thus $N=145$ deformation vectors $\vec{\Delta}_m$.
The units are in nm. The entry $\widetilde{A}^{tr}_q$ has been
multiplied by a factor $10$ to facilitate its visibility. The
stiffness matrix has the structure given in \eqref{DNA:stiff}. All
its entries are symmetric in $q$, except for the tilt-roll term
$\widetilde{A}^{tr}_q$ which is anti-symmetric. Blue dashed lines: Fits
of the data to Eqs.~\eqref{DNA:fitoxs} and \eqref{DNA:fitoxa}, with
fitting parameters given in Table~\ref{DNA:table_fitoxDNA}. (b) Plots
of $l_\text{B}$ and $l_\text{T}/2$ vs. $m$ the relative distance in
numbers of basepair-steps between the considered segments.  Green lines
are obtained from the stiffness matrix data using Eqs.~\eqref{DNA:lT}
and \eqref{DNA:lB}. The red line is the approximation \eqref{app:lB}.
In this case the difference between the two approximations for
$l_\text{B}$ is very small.
Black dashed lines are obtained by direct calculations of correlation
functions from simulations. The oscillatory behavior of the bending
persistence length stems from a light helicity of the traced contour.}
\label{fig:stiff_fourier} 
\end{figure}
%%%%%%%%%%%%%%%%%%%%%%%%%%%%%%%%%%%%%%%%%%%%%%%%%%%%%%%%%%%%%%%%%%%%

%%%%%%%%%%%%%%%%%%%%%%%%%%%%%%%%%%%%%%%%%%%%%%%%%%%%%%%%%%%%%%%%%
\begin{table}[b]
\caption{Summary of the stiffnesses in oxDNA2 (data in nm). $X_m$
are the fitting coefficients used in Eqs.~\eqref{DNA:fitoxs} and
\eqref{DNA:fitoxa}.  The two rightmost columns give the stiffnesses
at $q=0$ and $q=\Delta q$, as representatives of the long length scale
behavior (see Eqs.~\eqref{DNA:lTq0} and \eqref{DNA:lBq0}).  The last two
lines give the persistence lengths as obtained from Eqs.~\eqref{DNA:lT}
and \eqref{DNA:lB}. We give the local ($m=1$) value and the asymptotic
one ($m \to \infty$). All parameters are given in nm.}
\begin{ruledtabular}
\begin{tabular}{c|cccc|cc}
& $X_0$ & $X_1$ & $X_2$ & $X_3$ & $q=0$ &$q=\Delta q$ \\
%& H. mean\\
\hline
$\widetilde{A}^t_q$ & 54& 17& 4.0 & 1.1 & 76 & 69 \\ %&50 \\
$\widetilde{A}^r_q$ & 38& 2 & 0.8 & 0.2 & 41 & 40 \\ %&37  \\
$\widetilde{C}_q$   & 78& 22& 6.5 & 1.3 & 108& 98 \\ %&74 \\
$\widetilde{G}_q$   & 23& 6.0& 1.9& 0.4 & 31 & 28 \\ %&21 \\
\hline
$\widetilde{A}^{tr}_q$ &  
& -0.9&     &     & 0 &-0.5 \\ %&- \\
\hline
\hline
$l_\text{B}$   && 40 ($m=1$)&& 45 ($m\to \infty$)& \\
$l_\text{T}/2$ && 63 ($m=1$)&& 84 ($m\to \infty$)& \\
\end{tabular}
\end{ruledtabular}
\label{DNA:table_fitoxDNA}
\end{table}
%%%%%%%%%%%%%%%%%%%%%%%%%%%%%%%%%%%%%%%%%%%%%%%%%%%%%%%%%%%%%%%%%

%%%%%%%%%%%%%%%%%%%%%%%%%%%%%%%%%%%%%%%%%%%%%%%%%%%%%%%%%%%%%%%%%%%%  
\begin{figure}[t]
\includegraphics[width=0.52\textwidth]{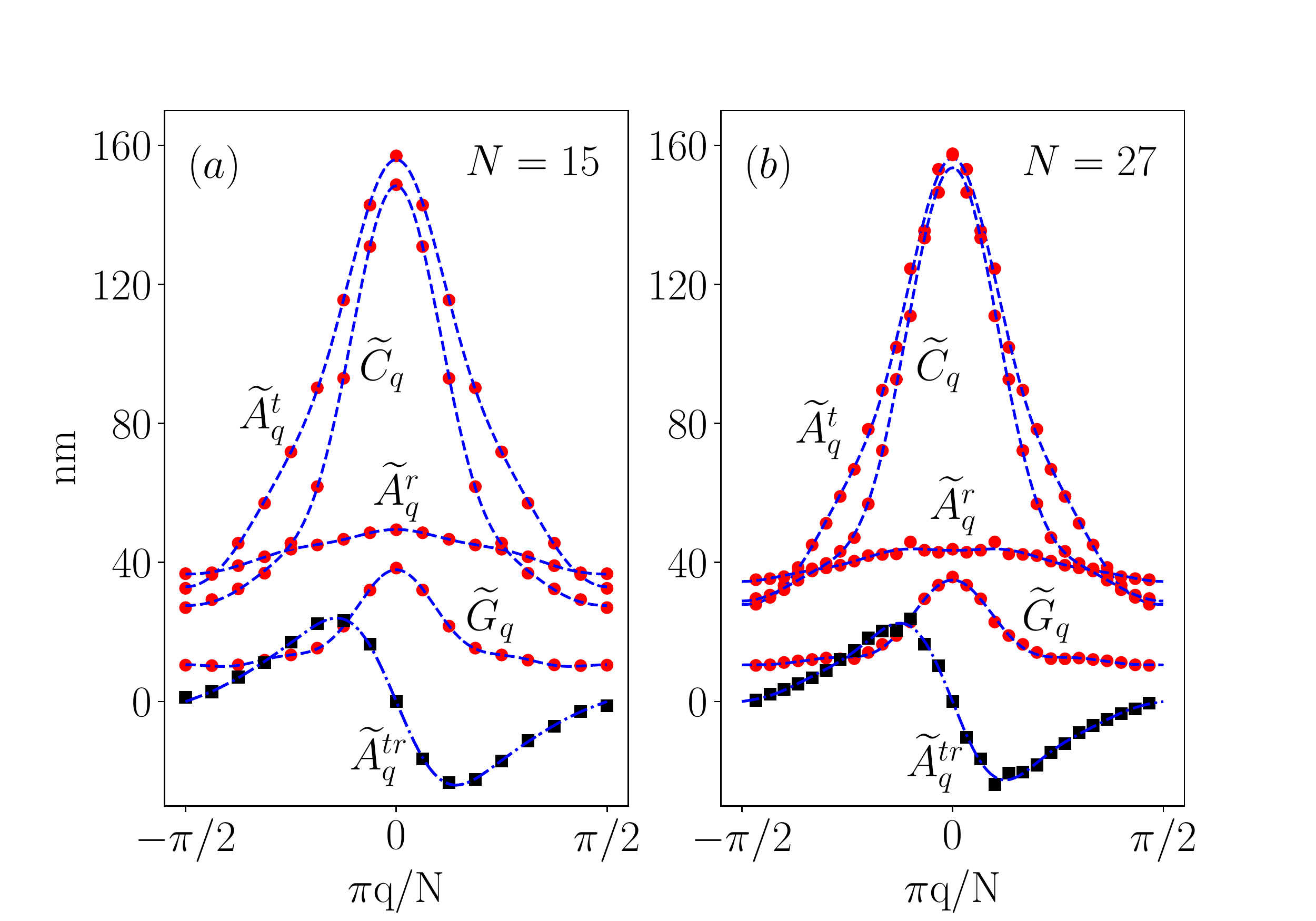}
\caption{Red dots and solid squares: Elements of the stiffness matrix
$\widetilde{M}_q$ as obtained from all-atom data for sequences of
length (a) $N=20$ (average of $9$ seq.) and (b) $N=32$ (average of
$3$ seq.). Dashed lines: fits of the forms \eqref{DNA:fitoxs} and
\eqref{DNA:fitoxa}.}
\label{fig:stiff_fourier-allatom} 
\end{figure}
%%%%%%%%%%%%%%%%%%%%%%%%%%%%%%%%%%%%%%%%%%%%%%%%%%%%%%%%%%%%%%%%%%%%

To quantify these effects the inverse Fourier transform of the data in
Fig.~\ref{fig:stiff_fourier}(a) was computed so as to obtain the couplings
in real space \footnote{Alternatively one can construct a global 
$3N \times 3N$ stiffness matrix  and extract the real space couplings from that
analysis. We verified that the results are the same.}. 
The Fourier series of the elements of the stiffness matrix
which are even or odd in $q$ are given by
\begin{eqnarray}
\widetilde{X}_q^\text{even} &=& \sum_m X_m \cos \frac{2m\pi q}{N},
\label{DNA:fitoxs}
\\
\widetilde{X}_q^\text{odd} &=& \sum_m X_m \sin \frac{2m\pi q}{N},
\label{DNA:fitoxa}
\end{eqnarray}
where $X_m$ are the real-space stiffness associated to couplings between
sites $n$ and $n+m$ \footnote{Note that the coupling \eqref{toy:Ktilde} of
the toy model can also be expresses as a Fourier series \eqref{DNA:fitoxs}
as follows $\tilde{K}_q = K+2K'+2K' \cos(2\pi q/N)$.}.

For the even terms we truncated the series to the first four
components, while in view of the uncertainties of the small
odd term $\widetilde{A}^{tr}_q$ we used a single term. The
best fits to the data are shown as dashed blue lines in
Fig.~\ref{fig:stiff_fourier}(a). Table~\ref{DNA:table_fitoxDNA}
gives the values of the corresponding coefficients $X_m$ resulting
from the fits. The coefficients decrease rapidly with $m$, but
there are significant off-site components for $\widetilde{C}_q$ and
$\widetilde{A}^r_q$, reflecting the strong $q$-dependence observed in
Fig.~\ref{fig:stiff_fourier}(a).
Twist and bend fluctuations are linked to the elements of the stiffness
matrix via the covariance matrix \eqref{DNA:cov}. Neglecting the small
contribution of $\widetilde{A}^{tr}_q$, and inverting $\widetilde{M}_q$
we get
\begin{equation}
\frac{a \langle |\widetilde{\Omega}_q|^2\rangle}{N} =
\frac{1}{ \widetilde{C}_q - \widetilde{G}_q^2/\widetilde{A}^r_q },
\label{oxDNA:twist}
\end{equation}
and
\begin{equation}
\frac{a \langle |\widetilde{\tau}_q|^2 +|\widetilde{\rho}_q|^2\rangle}{N}
= \frac{1}{\widetilde{A}^t_q} + \frac{1}{\widetilde{A}^r_q 
- \widetilde{G}_q^2/\widetilde{C}_q},
\label{oxDNA:bend}
\end{equation}
Inserting \eqref{oxDNA:twist} in \eqref{DNA:lT} we can estimate the
twist persistence length $l_T(m)$ from the stiffness data using the
truncated Fourier series as numerical estimates for $\widetilde{A}^r_q$,
$\widetilde{G}_q$ and $\widetilde{C}_q$. In a similar way inserting
\eqref{oxDNA:bend} into Eq.~\eqref{DNA:lB} allows us to calculate the
bending persistence length. The results of these calculations are shown
in Fig.~\ref{fig:stiff_fourier}(b) as solid green lines. The red solid
line is the approximation \eqref{app:lB} \footnote{We note that the $q=0$
component is $\vec{\Delta}_{q=0} = (\sum_n \tau_n, \sum_n \rho_n ,\sum_n
\Omega_n )$ The method introduced in \cite{skor17} derived asymptotic
stiffnesses using a covariance matrix obtained from the sums of $\tau_n$,
$\rho_n$ and $\Omega_n$ truncated to an increasing number of terms. Hence
the results reported in \cite{skor17} report the $q=0$ component of the
stiffness matrix.}. Dashed black lines show the direct calculations of
the bending persistence length as deduced from the decay length of the
respective correlation functions (\eqref{tw:correl} and \eqref{DNA:uu}).
While there is excellent overlap between dashed and solid lines for
$l_\text{T}$, some deviations of a few $nm$ are visible in $l_\text{B}$.
The overlap in $l_\text{T}$ was expected as \eqref{DNA:lT} is exact,
while both expressions \eqref{app:lB} and \eqref{DNA:lB} (red and
green lines in Fig.~\ref{fig:stiff_fourier}(b)) involve approximations.
Note also, that $l_\text{B}$ as deduced from the correlation function
exhibits damped oscillatory behavior stemming from a light helicity of
the used set of triads.

The last two lines of Table~\ref{DNA:table_fitoxDNA} give the local
($m=1$) and asymptotic ($m \to \infty$) values of the persistence
lengths as obtained from \eqref{DNA:lT} and \eqref{DNA:lB}. Both bending
and torsional persistence lengths are smaller at short distances as
compared to their asymptotic values, however the effect is modest
for $l_\text{B}$, while much stronger length-dependent variability is
observed in $l_\text{T}$. This can be understood from the elements of
the stiffness matrix. Torsional persistence is primarily determined
by $\widetilde{C}_q$ (Eq.~\eqref{oxDNA:twist}) which has a large $q$
dependence, causing strong length scale effects in $l_\text{T}$. On the
other side the bending stiffness is determined by the harmonic mean of
tilt and rescaled roll stiffnesses \eqref{oxDNA:bend}, which is dominated
by the softer roll component. The weak dependence of $\widetilde{A}^r_q$
on q in Fig.~\ref{fig:stiff_fourier}(a), indicating small off-site
roll-roll couplings, is the cause of the modest length scale dependence
of $l_\text{B}$.

\subsection{All-atom}

All-atom simulations of double stranded DNA of two different lengths were
performed. Details of setup, force fields, methodology and sequences
used can be found in Appendix~\ref{app:all_atom}. Tilt, roll and twist
variables were obtained from simulation data using an own implementation
of the algorithm underlying Curves+ \cite{lave09}. Subtracting the
averages we obtained the excess values $\vec{\Delta}_n^\intercal
=(\tau_n,\rho_n,\Omega_n)$. Local elasticity in all-atom models of DNA
is dependent on the type of base pairs, as opposed to the homogeneous
oxDNA model. Using the relation \eqref{DNA:cov} we derived an effective
stiffness matrix $\widetilde{M}_q$. The procedure builds up an equivalent
homogeneous model which shares the same covariance matrix as the
original data set by matching the second moments of the fluctuations
in Fourier space. For a system breaking translational invariance,
in general, the correlator $\langle \widetilde{\vec{\Delta}}_{q}
\widetilde{\vec{\Delta}}_{q'}^\dagger \rangle$ is non-zero also for
$q\neq q'$. In constructing the average stiffness matrix we ignore these
off-diagonal terms, which are expected to have weaker effect as the
system size grows, where effective translation invariance is recovered.

Figure~\ref{fig:stiff_fourier-allatom} shows the elements of
$\widetilde{M}_q$ in function of $y=\pi q/N$ as obtained from
this procedure (red dots and black squares). The lengths simulated
correspond to (a) $20$-mers and (b) $32$-mers, averaged over $10$
and $3$ different sequences, respectively.  Two nucleotides at each
end were removed from the analysis to  mitigate end effects. Hence
Fig.~\ref{fig:stiff_fourier-allatom} shows the Fourier transforms on
(a) $N=15$ and (b) $N=27$ data points.  Despite the difference in
length, the two sets exhibit quantitatively very similar stiffnesses.
The data share several common features with the oxDNA simulations of
Fig.~\ref{fig:stiff_fourier}: the tilt $\widetilde{A}^t_q$ and twist
$\widetilde{C}_q$ stiffnesses are strongly $q$-dependent, indicating
considerable contributions from off-site interactions. Just as for oxDNA
the roll stiffness $\widetilde{A}^r_q$ depends very weakly on $q$ and
again the only symmetric off-diagonal term of the stiffness matrix is the
twist-roll coupling $\widetilde{G}_q$.  Contrasting oxDNA in all-atom data
the tilt stiffness is larger than the twist stiffness $\widetilde{A}^t_q
> \widetilde{C}_q$ and their values are quantitatively much larger. In
addition the $q$-odd tilt-roll coupling $\widetilde{A}^{tr}_q$ is much
more prominent than in oxDNA.

%%%%%%%%%%%%%%%%%%%%%%%%%%%%%%%%%%%%%%%%%%%%%%%%%%%%%%%%%%%%%%%%%
\begin{table}[b]
\caption{All atom data for $20$-mers ($N=15$) and $32$-mers ($N=27$)
averaged over $10$ and $3$ different oligomers respectively.
All parameters are given in nm.}
\begin{ruledtabular}
\begin{tabular}{c|ccccc|cc}
N=15& $X_0$ & $X_1$ & $X_2$ & $X_3$ & 
  $X_4$ & $q=0$ & $q=\Delta q$\\
\hline
$\widetilde{A}^t_q$ &
82   & 56  & 11  & 5.8 & 1.3 & 156 & 130 \\
$\widetilde{A}^r_q$ &
43   & 5.9 & -0.4& 0.6 & 0.3 & 50  & 48  \\
$\widetilde{C}_q$   &
65   & 52  & 21  & 8.5 & 1.9 & 148 & 112   \\
$\widetilde{G}_q$   &
17   & 11  & 5.4 & 2.9 & 1.4 & 38  & 27   \\
\hline
$\widetilde{A}^{tr}_q$ &  
& -19& -8.4  & 0.3 & -0.6 & 0 &-21 \\
\hline
\hline
N=27& $X_0$ & $X_1$ & $X_2$ & $X_3$ & 
  $X_4$ & $q=0$ & $q=\Delta q$\\
\hline
$\widetilde{A}^t_q$ &
75   & 57  & 14  & 6.7 & 2.6 & 156 & 125 \\
$\widetilde{A}^r_q$ &
40   & 4.7 & -0.4& -0.2 & -0.5 & 43  & 44  \\
$\widetilde{C}_q$   &
67   & 53  & 23  & 9.3 & 1.4 & 154 & 116   \\
$\widetilde{G}_q$   &
17   & 9.3 & 5.0 & 2.9 & 1.0 & 35  & 25   \\
\hline
$\widetilde{A}^{tr}_q$ &
& -16& -8.9  & 0.4 & -1.2& 0 &-21 \\
\hline
\hline
$l_\text{B}$   && 42 ($m=1$)&& 61 ($m\to \infty$)& \\
$l_\text{T}/2$ && 43 ($m=1$)&& 125 ($m\to \infty$)& \\
\end{tabular}
\end{ruledtabular}
\label{DNA:table_fitAA}
\end{table}
%%%%%%%%%%%%%%%%%%%%%%%%%%%%%%%%%%%%%%%%%%%%%%%%%%%%%%%%%%%%%%%%%

Table~\ref{DNA:table_fitAA} shows the results of the fits of the elements
of $\widetilde{M}_q$ to Eqs.~\eqref{DNA:fitoxs} and \eqref{DNA:fitoxa}.
The coefficients $X_m$ decrease significantly with $m$, but more gradually
as compared to oxDNA, indicating more pronounced off-site interactions.
Overall, there is a only a small difference between the two data-sets,
which is indicative for weak finite size effects.  Using the coefficients
$X_m$ of the $N=27$ data set as representatives for the couplings of
a long DNA sequence we invoked \eqref{DNA:lT} and \eqref{DNA:lB} to
estimate the twist and bending persistence lengths.  Results are shown in
Fig.~\ref{fig:persl-allatom}(a). As in oxDNA $l_\text{T}$ has a strong
length scale dependence, while for $l_\text{B}$ this dependence in much
more modest.  The variability of $l_\text{T}$ across different length
scales is much larger in the all-atom data than in oxDNA. This is due
to the much stronger $q$-dependence of the stiffnesses of the former
as can be seen when comparing Fig.~\ref{fig:stiff_fourier-allatom}
to Fig.~\ref{fig:stiff_fourier}.  Interestingly, $l_\text{T}/2$
approaches an asymptotic value close to $130$~nm, which is not far from
the torsional stiffnesses ($120$~nm) measured in magnetic tweezers
\cite{lipf14}. This technique probes the torsional elasticity by
tracing the twist fluctuations of the ends of stretched DNA molecules
of several kilobases length. The recent atomistic simulation study
by Velasco-Berreleza et al. \cite{vela20} found a similarly strong
length-dependence of the torsional fluctuations, although their
asymptotic estimate indicates $l_\text{T}/2 \approx 90$~nm. We note
here that $l_\text{T}$ at all length scales is not only determined by
the twist stiffness $\widetilde{C}_q$, but also by other stiffnesses.
In oxDNA twist fluctuations are also influenced by $\widetilde{G}_q$
and $\widetilde{A}^{tr}_q$, see Eq.~\eqref{oxDNA:twist}. The relation
is even more elaborate if one includes the tilt-roll coupling
$\widetilde{A}^{tr}_q$, which is non-negligible in all atom data.

Figure~\ref{fig:persl-allatom}(b) shows our calculation of the
%  allosteric 
response of a DNA molecule to a generalized force imposed on
a certain basepair-step, as given by the integral \eqref{DNA:allostery}.
The generalized force $(f_\tau, f_\rho, f_\Omega)$ was tuned in order to
shift the average deformations ($\langle \tau_0 \rangle$, $\langle \rho_0
\rangle$ and $\langle \Omega_0 \rangle$) from zero to some finite angles
($20^o$, $25^o$ and $-20^o$ for tilt, roll and twist respectively).
Due to the presence of non-local couplings, neighboring steps are
expected to also be effected by this imposed force. The calculation
shows that the resulting shift in the average values decay very rapidly
to zero, which is the unperturbed value, with angles being negligibly
small already at $m=2$.  Although off-site couplings are capable of
carrying the effect of a local perturbation to distant flanking sites,
the characteristic decay length $l_\text{A}$ is quite small.  Why are
the twist and, to a more limited extent, the bending elasticity varying
so much with the length scale (Fig.~\ref{fig:persl-allatom}(a)), while
local pertubations (Fig.~\ref{fig:persl-allatom}(b)) decay so rapidly?
To understand this issue it is useful to go back to the toy model of
Section~\ref{sec:toy}.  At different length scales the elasticity is
governed by different stiffnesses ranging from $K_1$ to $K_{\infty}$,
where the asymptotic value is approached as $1/m$ for large $m$
(see Eq.~\eqref{Kexp}).  A local perturbation, on the contrary, decays
exponentially with a length linked to the relative difference between the
two local elastic constants $K_1$ and $K_2$, see Eq.~\eqref{toy:def_xi}.

%%%%%%%%%%%%%%%%%%%%%%%%%%%%%%%%%%%%%%%%%%%%%%%%%%%%%%%%%%%%%%%%%%%%  
\begin{figure}[t]
\includegraphics[width=0.46\textwidth]{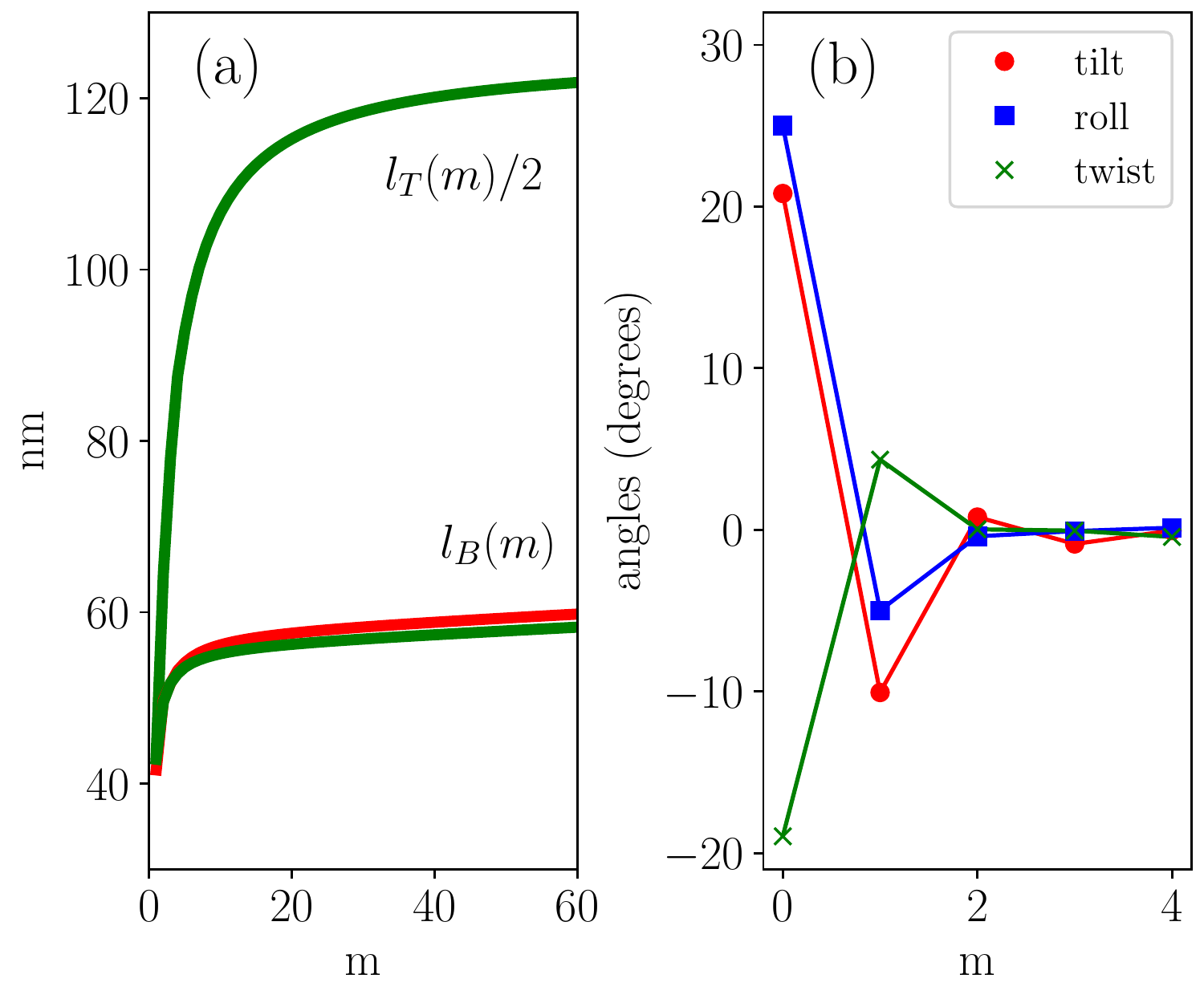}
\caption{(a) Estimated length scale dependence of the persistence
lengths as obtained from the analysis of the all-atom data in
Table~\ref{DNA:table_fitAA}. Assuming that these data are representatives
for the behavior of very long sequences, we used Eqs.~\eqref{DNA:lB}
and \eqref{DNA:lT} to calculate $l_\text{B}$ and $l_\text{T}$ 
(green lines). The red line is the approximation \eqref{app:lB}
for $l_\text{B}$.
(b) Calculation of the the propagation of pertubations induced by
generalized forces acting on the site $m=0$. This data is calculated with
Eq.~\eqref{DNA:allostery} using the data in Table~\ref{DNA:table_fitAA}.
Results are given in degrees (the quantities plotted are $180 \,
a\tau/\pi$, $180 \, a\rho/\pi$ and $180 \, a\Omega/\pi$).}
\label{fig:persl-allatom} 
\end{figure}
%%%%%%%%%%%%%%%%%%%%%%%%%%%%%%%%%%%%%%%%%%%%%%%%%%%%%%%%%%%%%%%%%%%%

%  \input{Sections/conclusion}
\section{Discussion}
\label{sec:conclusion}

In this paper we investigated the effects of interactions in DNA models
that extend beyond nearest-neighbors (off-site couplings). Our analysis
is based on the calculation of the stiffness matrix in momentum space
$\widetilde{M}_q$ for oxDNA and all-atom models. Both systems show very
similar behavior, which is presumably a consequence of the geometrical
structure of the double helix.  The set of matrices $\widetilde{M}_q$
encodes both the asymptotic long length scale stiffness $q= 0$ as well
as the short scale behavior obtained from harmonic means of the data. We
summarize here the main findings.

\subsection{General structure of the coupling matrices} 

Both oxDNA and all-atom data indicate that the general form of the
off-site coupling matrices can be understood from symmetry arguments,
generalizing those used to describe on-site interactions \cite{mark94}.
This symmetry requires the functional form of homogeneous models to
be invariant under reversal of the curvilinear coordinate, such that
that the first segment becomes the last and vice versa.  The resulting
generic form of $\widetilde{M}_q$ is given by Eq.~\eqref{DNA:stiff} and
contains terms which are either even or odd in $q$.  As odd terms vanish
in the limit $q \to 0$ they have a weak impact on the asymptotic length
scale elasticity, but they turn out to be more relevant at short length
scales. Our analysis confirms previous studies \cite{skor17} showing
that the twist-roll coupling $\widetilde{G}_q$ (even function of $q$)
is the dominant off-diagonal stiffness coefficient.

\subsection{Length dependence of persistence lengths}

Our analysis has shown that of the three rotational modes, tilt- ($\tau$)
and twist- ($\Omega$) exhibit significant off-site couplings. This
can be seen from the strong $q$ dependence of the respective momentum
space couplings ($\widetilde{A}^t_q$ and $\widetilde{C}_q$) as shown in
Figs.~\ref{fig:stiff_fourier}(a) and \ref{fig:stiff_fourier-allatom},
or equivalently in the appreciable real space coupling that extend up
the fourth neighbor in the case of the atomistic simulations (see
table~\ref{DNA:table_fitAA}). On the other hand, the remaining mode
roll ($\rho$) shows but modest off-site interactions, i.e.  a very weak
$q$-dependence of the momentum space couplings ($\widetilde{A}^r_q$). In
all cases the mode stiffness is softer locally and becomes increasingly
stiffer towards the asymptotic long range regime.  From the behavior
of these three modes one can understand the length dependence of the
twist and bending persistence length.  The twist persistence length
$l_\text{T}$ is fully determined by the behavior of the twist degree
if freedom and therefore mirrors its strong length dependence (see
Figure \ref{fig:stiff_fourier}(b)), which is in agreement with previous
studies \cite{noy12}. In the case of oxDNA2, manifests in an about $35\%$
increase in stiffness from the local to the asymptotic elasticity. The
bending persistence length $l_\text{B}$ is determined by the harmonic
mean of the stiffnesses governing the fluctuations of the two bending
modes $\tau$ and $\rho$, which is dominated by the softer $\rho$ mode
(see Eqs.~\eqref{DNA:lBq0} and \eqref{oxDNA:bend}). Accordingly, the
weak length dependence of this mode translates into a likewise behavior
of the bending persistence length.  We observed similar effects in the
all atom data, although the difference in torsional elasticity at short
and very long length scales is much larger in that case, as illustrated
in Fig.~\ref{fig:persl-allatom}.  This strong length scale dependence of
the torsional elasticity can potentially explain the divergence between
estimates obtained with different experimental methods~\cite{vela20}.
Studies that employ local probing methods 
find systematically lower stiffnesses as compared to studies in which
larger length scales are considered, as is the case for magnetic tweezers
%  (a table with several different estimates are given in the supplemental
%  of Ref.~\cite{nomi17}).
(for a list of different estimates and methods used see supplemental
of Ref.~\cite{nomi17}).

%  \subsection{Distal allostery}
\subsection{Local perturbations}

%  A topic of recent interest is the study of allosteric interactions in DNA.
%  Here one distinguishes between proximal and distal allostery. Proximal
%  allostery typically involves the binding of small molecules in the DNA
%  minor groove which alter the corresponding major groove binding site
%  affinity for a protein, see for instance the discussion in \cite{chen09}
%  and \cite{drsa14}. Distal allostery involves perturbations occurring
%  outside the binding site \cite{kim13}.  
Our model predicts that local DNA deformations such as an imposed
bending or twist angle at a given site induces structural changes of
the flanking sites up to some characteristic distance. This distance
depends both on the magnitude of the off-diagonal couplings and the range
of the interactions. For the analyzed models we find that the effect is
rather modest, with the perturbation involving just three flanking sites.
Experiments analyzing DNA-proteins interactions have highlighted a few
cases of distal allosteric effects \cite{kim13,rose20}, where the binding
of a protein at a given site increases the binding affinity to a second
protein. This distance is of about $15-20$ nucleotides.  A more common
phenomenon is that of proximal allostery, which involves the binding of
small molecules in the DNA minor groove altering the corresponding major
groove binding site affinity for a protein (see for example the discussion
in \cite{chen09} and \cite{drsa14}). Our analysis indicates that, within
linear elasticity, distal allostery is rather modest as compared to the
distal effects seen in these experiments \cite{kim13,rose20}.  This short
perturbation range was obtained from the average elastic behavior of the
considered sequences. It remains to be seen if some specific sequences can
exhibit a much more pronounced effect. Beyond that, it is likely that,
in order to fully account for the experimentally observed allostery,
one would need to go beyond linear elasticity, see e.g.\ \cite{sing18}.

\medskip

To conclude, we remark that, while we restricted our analysis to
rotational deformations, it could be extended to include translational
inter-basepair degrees of freedom.  In our opinion an accurate account
of off-site interactions is very useful for a deeper understanding of
DNA elasticity and how the local behavior crosses over to long scale
asymptotic properties.

\appendix

%%%%%%%%%%%%%%%%%%%%%%%%%%%%%%%%%%%%%%%%%%%%%%%%%%%%%%%%%%%%%%%%%%%%  
\begin{figure}[t]
\includegraphics[width=0.42\textwidth]{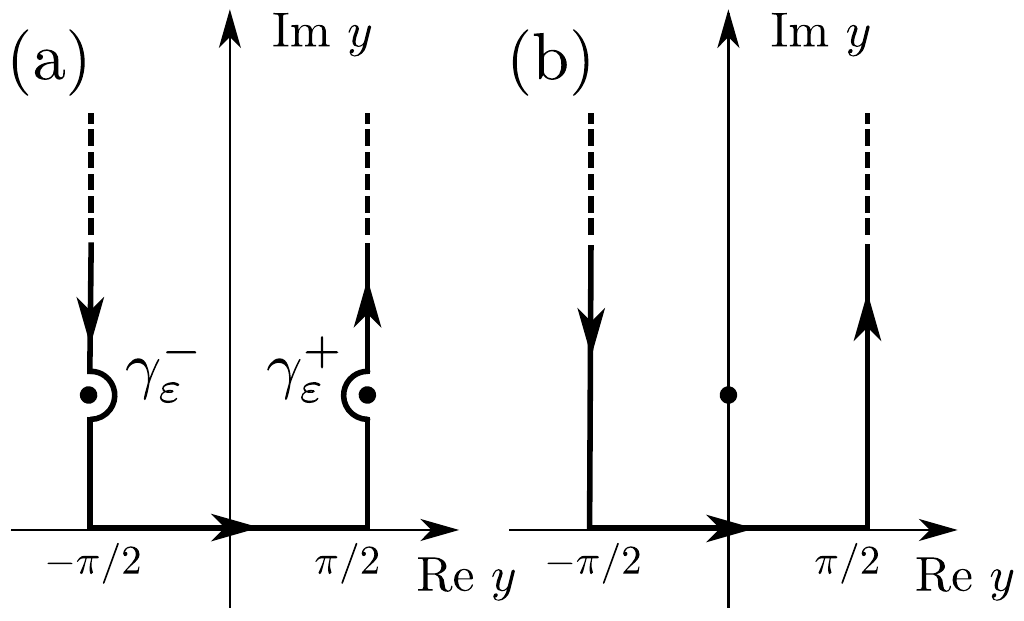}
\caption{Integration contours in the complex $y$-plane used for the
evaluation of the integral \eqref{app:allostery}. The two cases 
correspond to: (a) $K'>0$ and (b) $-K/4 < K' < 0$.}
\label{fig:contour}
\end{figure}
%%%%%%%%%%%%%%%%%%%%%%%%%%%%%%%%%%%%%%%%%%%%%%%%%%%%%%%%%%%%%%%%%%%%  

\section{Decay of local perturbation}
\label{app:allostery}

We give here further details about the calculation of the integral in
Eq.~\eqref{toy:allostery}
\begin{eqnarray}
I &=& \frac{\beta f}{\pi} 
\int_{-\pi/2}^{\pi/2} \frac{e^{2iym} \, dy}{K+4K' \cos^2 y}.
\label{app:allostery}
\end{eqnarray}
As mentioned earlier stability of the model requires that either $K'
>0$ or $-K/4 < K' <0$. We will discuss these two cases separately.

\subsection{$K' > 0$}

In this case the integrand has two simple poles in $y = \pm \pi/2
+ i \alpha$ with $\alpha >0$ the solution of $\cosh^2 \alpha =
K/4K'$.  We extend the integration over the contour indicated in
Fig.~\ref{fig:contour}(a), which is closed at infinity. The integral in
this domain does not enclose any singularities hence it vanishes. The
integrals along the two vertical lines cancel each other, due to symmetry,
so one is left with
\begin{eqnarray}
I + \frac{\beta f}{\pi} 
\int_{\gamma_\varepsilon^+ \cup \gamma_\varepsilon^-} 
\frac{e^{2iym} \, dy}{K+4K' \cos^2 y} &=& 0,
\label{app:allostery2}
\end{eqnarray}
where 
\begin{eqnarray}
\gamma_\varepsilon^{\pm} (\phi) = i \alpha \pm \frac{\pi}{2} 
+ \varepsilon e^{-i\phi},
\end{eqnarray}
are the two small half-circles around the two poles. The integrations in
these two domains pick up contributions from the poles and directly yield
the expression \eqref{toy:allostery}. In particular, the oscillating
behavior stems from the fact that the poles are in $\pm \pi/2$, which
leads to the appearance of a factor $\exp(\pm im\pi) = (-1)^m$.  The
associated decay length is then simply given by $l_\text{A} = 1/2\alpha$.

\subsection{$-K/4 < K' < 0$}

In this case the integrand has a simple pole in $y=i \alpha$ with $\alpha
>0$ the solution of the equation $\cosh^2 \alpha = K/4|K'|$. We extend
the integration to the domain shown in Fig.~\ref{fig:contour}(b). The
integration picks up the residue from the pole along the imaginary axis.
Thus, one can again obtain $I$.  Note that, as the pole is purely
imaginary, there are no oscillations, but a pure exponential decay.

More complicated integrands will eventually contain several poles,
giving rise to a sum of exponentials. The dominant contribution will be
given by the pole in the semi-infinite strip $-\pi/2 \leq \text{Re}(y)
\leq \pi/2$, $\text{Im}(z) >0$ which is closest to the real axis.

\section{Bending persistence length}
\label{app:rot_matrices}

The rotation operator mapping the triad $( \widehat{{\vec f}}_k\widehat{{\vec
v}}_k  \widehat{{\vec u}}_k)$ into $( \widehat{{\vec
f}}_{k+1} \widehat{{\vec v}}_{k+1}  \widehat{{\vec u}}_{k+1})$ can
be expressed as
\begin{equation}
    {\cal R}_k = 
\widehat{{\vec f}}_{k+1} \otimes \widehat{{\vec f}}_k +
\widehat{{\vec v}}_{k+1} \otimes \widehat{{\vec v}}_k +
\widehat{{\vec u}}_{k+1} \otimes \widehat{{\vec u}}_k.
\label{app:defrot}
\end{equation}
Here $\otimes$ denotes the tensor product, which transforms
a generic vector $\vec a$ as follows
\begin{equation}
\left(  \vec{u} \otimes \vec{v} \right) \vec{a} = 
(\vec{a} \cdot \vec{v}) \vec{u}.
\end{equation}
From \eqref{app:defrot} it follows that
${\cal R}_k \widehat{{\vec f}}_k = \widehat{{\vec f}}_{k+1}$,
${\cal R}_k \widehat{{\vec v}}_k = \widehat{{\vec v}}_{k+1}$ and
${\cal R}_k \widehat{{\vec u}}_k = \widehat{{\vec u}}_{k+1}$.
An alternative ``axis-angle" representation uses a unit vector
$\widehat{\boldsymbol\gamma}$ 
as rotation axis and
a rotation angle $\theta$. For a counterclockwise rotation around
$\widehat{\boldsymbol\gamma}$ this representation takes the form
\begin{equation}
    {\cal R} = \cos \theta \left(1 -  
\widehat{\boldsymbol\gamma} \otimes \widehat{\boldsymbol\gamma}
\right) + \sin \theta \, (\boldsymbol{\epsilon\widehat{\gamma}}) + 
\widehat{\boldsymbol\gamma} \otimes \widehat{\boldsymbol\gamma},
\label{app:rot_tensor}
\end{equation}
where 
\begin{equation}
  ( \epsilon \vec u ) \vec{a} = \vec{u} \times \vec{a}. 
\end{equation}
One can easily verify from \eqref{app:rot_tensor} that ${\cal
R} \widehat{\boldsymbol\gamma} = \widehat{\boldsymbol\gamma}$
and that for any unit vector $\widehat{\vec a}$ orthogonal
to $\widehat{\boldsymbol\gamma}$ the following relations hold:
(a) $\widehat{\boldsymbol\gamma} \cdot {\cal R}\widehat{\vec a} =
0$ and (b) $\widehat{\vec a} \cdot {\cal R} \widehat{\vec a} = \cos
\theta$. This shows that the rotated vector ${\cal R} \widehat{\vec a}$
is orthogonal to the rotation axis and that it forms an angle $\theta$
with $\widehat{\vec a}$.
As mentioned in the main text tilt, roll and twist are the components
of the Euler vector with respect to the local triad
\begin{equation}
    \vec{\Theta} = a \tau \widehat{\vec f} + 
    a \rho \widehat{\vec v} +
    a (\Omega + \omega_0 ) \widehat{\vec u},
\label{app:defTheta2}
\end{equation}
where its length $\Theta \equiv |\vec\Theta|$ gives the rotation angle. It
is convenient to define
\begin{equation}
\text{t} \equiv a\tau/\Theta, \qquad 
\text{r} \equiv a\rho/\Theta, \qquad
\text{w} \equiv a(\Omega+\omega_0)/\Theta,
\label{app:deftrw}
\end{equation}
for which $\text{t}^2 + \text{r}^2 + \text{w}^2 =1$ holds.
Using \eqref{app:rot_tensor} with $\widehat{\boldsymbol\gamma} =
\vec\Theta_k/\Theta_k$ and $\theta=\Theta_k$ and \eqref{app:defTheta2}
one finds
\begin{eqnarray}
\widehat{\vec u}_{k+1} &=& {\cal R}_{k} \widehat{\vec u}_{k} = 
\left[ \cos \Theta_{k} + (1-\cos \Theta_{k}) 
\text{w}_k^2
\right] \widehat{\vec u}_{k}
\nonumber \\
&+& \left[ (1-\cos \Theta_k) \text{t}_k \text{w}_k
+ \sin \Theta_k \text{r}_k
\right] \widehat{\vec f}_{k}
\nonumber \\
&+& \left[ 
(1-\cos \Theta_k) 
\text{r}_k \text{w}_k
- \sin \Theta_k \text{t}_k
\right] \widehat{\vec v}_{k}.
\label{app:prodkk+1}
\end{eqnarray}
%  In an analogous manner one can work out the other two products
%  $\widehat{\vec f}_{k+1} = {\cal R}_{k} \widehat{\vec f}_{k}$ and
%  $\widehat{\vec v}_{k+1} = {\cal R}_{k} \widehat{\vec v}_{k}$.
%  The (\ref{app:prodkk+1}) and the last two relations can
%  be written in a compact matrix form as follows
This relation, together with the two relations obtained from
$\widehat{\vec f}_{k+1} = {\cal R}_{k} \widehat{\vec f}_{k}$ and
$\widehat{\vec v}_{k+1} = {\cal R}_{k} \widehat{\vec v}_{k}$ can be
cast in a matrix product form as
\begin{equation}
\begin{pmatrix}
\widehat{\vec f}_{k+1} \\ 
\widehat{\vec v}_{k+1} \\ 
\widehat{\vec u}_{k+1}
\end{pmatrix}
    = {\vec R}_k
\begin{pmatrix}
\widehat{\vec f}_{k} \\ 
\widehat{\vec v}_{k} \\ 
\widehat{\vec u}_{k}
\end{pmatrix}.
\label{app:mat_mult}
\end{equation}
The $3 \times 3$ matrix ${\vec R}_k$ is given by
\begin{widetext}
\begin{equation}
{\vec R} = 
\begin{pmatrix}
\cos \Theta + (1-\cos \Theta) \, \text{t}^2
&& (1 - \cos \Theta) \text{t} \, \text{r} +\sin \Theta \, \text{w}
&& (1 - \cos \Theta) \text{t} \, \text{w} -\sin \Theta \, \text{r}\\
   (1 - \cos \Theta) \text{t} \, \text{r} -\sin \Theta \, \text{w}
&&  \cos \Theta +(1 -  \cos \Theta) \, \text{r}^2
&& (1 - \cos \Theta) \text{r} \, \text{w}+\sin \Theta \, \text{t} \\
   (1 - \cos \Theta) \text{t} \, \text{w}+\sin \Theta \, \text{r}
&& (1 - \cos \Theta) \text{r} \, \text{w}-\sin \Theta \, \text{t}  
&&  \cos \Theta +(1 -  \cos \Theta) \, \text{w}^2
\end{pmatrix},
\label{app:rot_mat}
\end{equation}
\end{widetext}
where for simplicity we dropped the index $k$.
Setting $k=m-1$, Eq.~\eqref{app:prodkk+1} reads
\begin{equation}
\widehat{\vec u}_{m} = 
\left( \vec R_{m-1} \right)_{31} \widehat{\vec f}_{m-1} + 
\left( \vec R_{m-1} \right)_{32} \widehat{\vec v}_{m-1} + 
\left( \vec R_{m-1} \right)_{33} \widehat{\vec u}_{m-1} ,
\end{equation}
a relation that can be iterated further using $\widehat{\vec f}_{m-1}
= {\cal R}_{m-2} \widehat{\vec f}_{m-2}$, $\widehat{\vec v}_{m-1}
= {\cal R}_{m-2} \widehat{\vec v}_{m-2}$, $\widehat{\vec u}_{m-1} =
{\cal R}_{m-2} \widehat{\vec u}_{m-2}$ and similar relations for $m-2$,
$m-3$\ldots.  In this way one expresses $\widehat{\vec u}_{m}$ as a linear
combination of $\{\widehat{\vec f}_0,\widehat{\vec v}_0, \widehat{\vec
u}_0\}$ with coefficients given as products of rotation matrices
\eqref{app:rot_mat}. The tangent-tangent correlator \eqref{DNA:uu}
then becomes the element $33$ of the product of these matrices
\begin{eqnarray}
{\cal C}_B (m) &=& 
\left\langle \widehat{\vec u}_{0} \cdot 
\widehat{\vec u}_{m} \right\rangle =
\left\langle 
{\vec R}_{m-1} \ldots {\vec R}_1 {\vec R}_0 
\right\rangle_{33}.
\label{app:CB}
\end{eqnarray}

Next, we develop two approximations for the calculation of ${\cal
C}_B (m)$. The first one assumes that the rotation angle $\Theta$
to be infinitesimal. The second one, which is a better approximation,
relies on the fact that for DNA the rotation from one basepair attached
triad to the next is dominated by the intrinsic twist component.

\subsection{Infinitesimal rotations}

We consider the limit $\Theta \to 0$ and develop $\cos \Theta$ and 
$\sin \Theta$ in \eqref{app:rot_mat} to lowest order in $\Theta$.
Formally, this can also be considered as the continuum limit $a \to 0$,
which gives to lowest order (using \eqref{app:deftrw})
\begin{eqnarray}
{\vec R}_{33} &=& 1 - \frac{\Theta^2}{2} (1-\text{w}^2) =
1 - \frac{\Theta^2}{2} (\text{t}^2 +\text{r}^2) 
\nonumber \\
&=&
1 - \frac{a^2}{2} (\tau^2+\rho^2).
%  {\vec R}_{13} &\approx& -{\vec R}_{31} \approx a \rho \\
%  {\vec R}_{23} &\approx& -{\vec R}_{32} \approx -a \tau \\
%  {\vec R}_{11} &\approx&  {\vec R}_{22} \approx 1 + {\cal O}(a^2) \\
%  {\vec R}_{12} &\approx&  -{\vec R}_{21} \approx {\cal O}(a) 
\end{eqnarray}
Likewise,
${\vec R}_{13} \approx -{\vec R}_{31} \approx -a \rho$,
${\vec R}_{23} \approx -{\vec R}_{32} \approx a \tau$ and similar
expressions for the other elements.
We consider next the product between two rotation matrices 
to lowest order in $a$. For instance, for the element 13 we get
\begin{eqnarray}
\left( {\vec R}_1 {\vec R}_0 \right)_{13} &=&
\left( {\vec R}_1 \right)_{11} 
\left( {\vec R}_0 \right)_{13} +
\left( {\vec R}_1 \right)_{12} 
\left( {\vec R}_0 \right)_{23} +
\nonumber\\
&&
\left( {\vec R}_1 \right)_{13} 
\left( {\vec R}_0 \right)_{33} 
= -a(\rho_{1}+\rho_{0}) + {\cal O}(a^2). 
\nonumber\\
\label{app:RR13}
\end{eqnarray}
We notice that, when calculating this product, we can set $({\vec
R}_{1})_{11}=1$ and $({\vec R}_{1})_{12}=0$ as their higher order
corrections in $a$ do not contribute to the lowest order in $a$ to the
end result in \eqref{app:RR13}. Analogously, when computing $({\vec
R}_{1} {\vec R}_0 )_{23}$ we can set $({\vec R}_{1})_{21}=0$ and $({\vec
R}_{1})_{22}=1$.
Summarizing, if one is interested in the $33$ entry of the product of
rotation matrices as in \eqref{app:CB} to lowest order in $a$, it is
sufficient to approximate a rotation matrix as
\begin{equation}
{\vec R}_n =
\begin{pmatrix}
1 && 0 &&-a\rho_n\\
0 && 1 && a\tau_n\\
 a\rho_n &&-a\tau_n && 1-\frac{a^2}{2} (\tau_n^2+\rho_n^2)
\end{pmatrix}.
\label{app:approx_Rn}
\end{equation}
The product of two such matrices (again to lowest order in $a$) gives
\begin{equation}
{\vec R}_1 {\vec R}_0 =
\begin{pmatrix}
1 && 0 && -a(\rho_1+\rho_0) \\
0 && 1 &&  a(\tau_1+\tau_0) \\
a(\rho_1+\rho_0) && -a(\tau_1+\tau_0) &&  X_{0,1}\\
\end{pmatrix},
\end{equation}
where we defined
\begin{eqnarray}
X_{0,1} &=& 
\left[ 1 - \frac{a^2}{2} \left( \tau_{1}^2 + \rho_{1}^2 \right)\right]
\left[ 1 - \frac{a^2}{2} \left( \tau_{0}^2 + \rho_{0}^2 \right)\right]
\nonumber\\
&&-a^2 \tau_0 \tau_{1}
-a^2 \rho_0 \rho_{1} 
\nonumber\\
&=& 
1 - \frac{a^2}{2} \left[
\left( \tau_0 + \tau_{1} \right)^2 +
\left( \rho_0 + \rho_{1} \right)^2 
\right]+ {\cal O}(a^4).
\nonumber\\
\label{app:Xnnp1}
\end{eqnarray}
In conclusion, the product yields again a matrix of the form
\eqref{app:approx_Rn} with tilt and roll given as the sum of the tilt and
roll of the two matrices. This can be generalized to the product of $m$
matrices
\begin{equation}
\left(
{\vec R}_{m-1} \ldots {\vec R}_{1} {\vec R}_{0}
\right)_{33} =
1 - \frac{a^2}{2} \left[
\left( \sum_{k=0}^{m-1} \tau_{k} \right)^2 +
\left( \sum_{k=0}^{m-1} \rho_{k} \right)^2 
\right].
\end{equation}
Combining this last result and Eq.~\eqref{DNA:deflB} we get
\begin{equation}
\frac{1}{l_\text{B}} = \frac{a}{2m} 
\left\langle \left( \sum_{k=0}^{m-1} \tau_{k} \right)^2 +
\left( \sum_{k=0}^{m-1} \rho_{k} \right)^2 \right\rangle,
\label{app:eqlB}
\end{equation}
which, as done for the torsional persistence length \eqref{DNA:lT}, 
in the limit $N \to \infty$ can be written as
\begin{equation}
\frac{1}{l_\text{B}}
= \frac{a}{\pi m} \int_{-\pi/2}^{\pi/2} \frac{\sin^2 my}{\sin^2 y}
\, \frac{ \left\langle |\widetilde{\tau}_{q} |^2 +
|\widetilde{\rho}_{q} |^2 \right\rangle}{N} \, dy,
\label{app:lB}
\end{equation}
where as in the main text $y = \pi q/N$.

\subsection{Intrinsic twist dominance}

An improved approximation scheme uses the fact that the rotation is
dominated by the intrinsic twist component. Indeed, in DNA one has
$\omega_0 \gg |\Omega|$, $|\tau|$, $|\rho|$, where the difference is
typically one order of magnitude. In degrees (note that $a\tau$, $a\rho$,
$a\Omega$ are otherwise given in radians), the intrinsic twist angle is
$a\omega_0 \approx 34^\circ$, while the other angles are a few degrees.
This suggests that one can decompose
\begin{equation}
{\vec R}_{n} = {\vec S} \widehat{\vec R}_{n},
\label{eq:split_Rn}
\end{equation}
as the product of two rotations where $\widehat{\vec R}_{n}$
is small and ${\vec S}$ a pure
twist rotation of magnitude $a\omega_0$. Setting
$\text{t}=\text{r}=0$, $\text{w}=1$ and $\Theta = a \omega_0$ in
\eqref{app:rot_mat} we have
\begin{equation}
{\vec S} = 
\begin{pmatrix}
 \cos(a{\omega}_0) &  \sin(a{\omega}_0) & 0 \\
-\sin(a{\omega}_0) &  \cos(a{\omega}_0) & 0 \\
   0 &    0 & 1 \\
\end{pmatrix}.
\label{app:defS}
\end{equation}
The product of two consecutive rotation matrices is
\begin{equation}
{\vec R}_{1}{\vec R}_{0} =  
{\vec S}^2
\left( {\vec S}^{-1} \widehat{\vec R}_{1} {\vec S} \right)
\widehat{\vec R}_{0} =
{\vec S}^2 {\vec R}^*_{1} {\vec R}^*_{0}, 
\label{eq:split_RR}
\end{equation}
where we defined
\begin{equation}
{\vec R}^*_{n} \equiv 
\left( {\vec S}^{-1}  \right)^n
\widehat{\vec R}_{n} {\vec S}^n = 
\left( {\vec S}^{-1} \right)^{n+1} 
{\vec R}_{n} {\vec S}^n .
\label{app:define_R*}
\end{equation}
%  Equation~\eqref{eq:split_RR} can be easily generalized to the product
%  of $m$ matrices:
For the product of $m$ matrices we get
\begin{equation}
{\vec R}_{m-1} \ldots {\vec R}_{1} {\vec R}_{0} =  
{\vec S}^{m} {\vec R}^*_{m-1} \ldots {\vec R}^*_{1} {\vec R}^*_{0}. 
\label{app:prodR*m}
\end{equation}
Taking the thermal average of the $33$ component of the two sides of the 
previous equation we find
\begin{equation}
{\cal C}_\text{B} (m) = 
\left\langle {\vec R}_{m-1} \ldots {\vec R}_{1} {\vec R}_{0}
\right\rangle_{33} = \left\langle {\vec R}^*_{m-1} \ldots {\vec R}^*_{1}
{\vec R}^*_{0} \right\rangle_{33},
\label{app:33R*}
\end{equation}
where we used $({\vec S}^{m})_{3k} =\delta_{3k}$.
To calculate the bending persistence length we will
be using the right hand side of \eqref{app:33R*}. 
%  We note that
%  \begin{equation}
%  {\vec R}^*_{n} = {\vec A}^T_{n} {\vec R}_n {\vec A}_{n-1}
%  \label{app:getR*}
%  \end{equation}
%  with
%  \begin{equation}
%  {\vec A}_n = {\vec S}_n {\vec S}_{n-1} \ldots {\vec S}_{0} = 
%  \begin{pmatrix}
%   c_n &  s_n & 0 \\
%  -s_n &  c_n & 0 \\
%    0  &   0  & 1 \\
%  \end{pmatrix}
%  \end{equation}
%  and
%  \begin{eqnarray}
%  c_n &\equiv& \cos \left[ (n+1)a \omega_0 \right] \\
%  s_n &\equiv& \sin \left[ (n+1)a \omega_0 \right]
%  \end{eqnarray}
%  Intrinsic twist dominance means that $\widehat{\vec v}$, the unit
%  vector giving the rotation axis in \eqref{app:rot_mat}, has elements
%  $v_3 \approx 1 \gg |v_1|, |v_2|$. 
Intrinsic twist dominance implies that in \eqref{app:rot_mat} $\text{w}
\approx 1$ and $|\text{t}|, |\text{r}| \ll 1$ and $\Theta \approx a
\omega_0$. We can use the approximations
\begin{equation}
\text{w} = \sqrt{1 - \text{t}^2 - \text{r}^2} 
\approx 1 - \frac{\text{t}^2 +\text{r}^2}{2}
= 1 + {\cal O}(\text{t}^2 , \text{r}^2),
\end{equation}
%  The rotation angle in \eqref{app:rot_mat} is approximated as 
%  \begin{equation}
%  \Theta = a {\omega}_0 + {\cal O}(\text{t}^2 , \text{r}^2)
%  \end{equation}
and $\Theta = a {\omega}_0 + {\cal O}(\text{t}^2 , \text{r}^2)$.
This implies that \eqref{app:rot_mat} to lowest orders in $\text{t}$ and $\text{r}$ 
becomes
%  This gives the following form to lowest orders in $\text{t}$ and $\text{r}$
%  \begin{widetext}
%  \begin{equation}
%  \R =
%  \begin{pmatrix}
%  c_0   &&& -s_0  &&&{\displaystyle\frac{(1-c_0) \tau +s_0 \rho}{\omega_0}}\\
%  %  \\
%  s_0   &&&  c_0  &&&{\displaystyle\frac{(1-c_0) \rho -s_0 \tau}{\omega_0}}\\
%  %  \\
%  {\displaystyle\frac{(1-c_0) \tau - s_0 \rho}{\omega_0}} &&& 
%  {\displaystyle\frac{(1-c_0) \rho + s_0 \tau}{\omega_0}} &&&
%  1- (1-c_0) {\displaystyle\frac{\tau^2+\rho^2}{\omega_0^2}}\\
%  \end{pmatrix}
%  \label{app:mat_app2}
%  \end{equation}
\begin{widetext}
\begin{equation}
{\vec R} =
\begin{pmatrix}
 \cos(a \omega_0) &&&& \sin(a \omega_0) &&&&  (1-\cos(a \omega_0))t -\sin(a \omega_0) r\\
-\sin(a \omega_0) &&&& \cos(a \omega_0) &&&&  (1-\cos(a \omega_0))r +\sin(a \omega_0) t\\
 (1-\cos(a \omega_0))t + \sin(a \omega_0) r &&&& 
 (1-\cos(a \omega_0))r - \sin(a \omega_0) t &&&& 
1 - (1-\cos(a \omega_0))(t^2+r^2)\\
\end{pmatrix}.
\label{app:rot_twist-dom}
\end{equation}
\end{widetext}
%  \begin{equation}
%  \R =
%  \begin{pmatrix}
%  c_0   &&& -s_0  &&&{\displaystyle\frac{(1-c_0) \tau +s_0 \rho}{\omega_0}}\\
%  %  \\
%  s_0   &&&  c_0  &&&{\displaystyle\frac{(1-c_0) \rho -s_0 \tau}{\omega_0}}\\
%  %  \\
%  {\displaystyle\frac{(1-c_0) \tau - s_0 \rho}{\omega_0}} &&& 
%  {\displaystyle\frac{(1-c_0) \rho + s_0 \tau}{\omega_0}} &&&
%  1- (1-c_0) {\displaystyle\frac{\tau^2+\rho^2}{\omega_0^2}}\\
%  \end{pmatrix}
%  \label{app:mat_app2}
%  \end{equation}
Note that taking $a \to 0$ one recovers the infinitesimal form
\eqref{app:approx_Rn}. As in that case, we can ignore terms dependent on
$\tau$, $\rho$ (t and r) in the upper $2 \times 2$ block as these will
not contribute to the bending persistence length to significant order.
%  Using the approximation \eqref{app:mat_app2} we can now work out
%  \eqref{app:getR*} to obtain $\R^*$. First we compute:
%  \begin{equation}
%  {\vec A}^T_n {\vec R}_n
%  = \begin{pmatrix}
%  c_{n}  & s_{n} &&  c_{n-1} {\vec R}_{13} - s_{n-1} {\vec R}_{23} \\
%  -s_{n} & c_{n} &&  s_{n-1} {\vec R}_{13} + c_{n-1} {\vec R}_{23} \\
%  {\vec R}_{31} & {\vec R}_{32} && {\vec R}_{33} \\
%  \end{pmatrix}
%  \end{equation}
%  where $\R_{13}$, $\R_{23}$ \ldots indicate the corresponding entries
%  of the matrix $\R_{n}$ (we have omitted the index $n$ to simplify the
%  notation). 
%  The multiplication by the matrix $A_{n-1}$ gives
%  \begin{equation}
%  \R^*_n = A_{n-1} \R_n A_{n}^T = 
%  \begin{pmatrix}
%  1 && 0 &&  c_{n-1} \R_{13} - s_{n-1} \R_{23} \\
%  0 && 1 &&  s_{n-1} \R_{13} + c_{n-1} \R_{23} \\
%   c_{n} \R_{31} - s_{n} \R_{32} &&
%   s_{n} \R_{31} + c_{n} \R_{32} && \R_{33} \\
%  \end{pmatrix}
%  \label{app:endR*}
%  \end{equation}
%  \end{widetext}
Next, we calculate $\vec{R}_n^*$ using the above form of $\vec{R}_n$
\eqref{app:rot_twist-dom} and Eq.~\eqref{app:define_R*}. The matrices
${\vec S}^n$ and $({\vec S}^{-1})^{n+1}$ have a block-diagonal form
as \eqref{app:defS} and correspond to a counterclockwise twist rotation
of an angle $n a \omega_0$ and a clockwise twist rotation of an angle
$(n+1) a \omega_0$, respectively.
%  The top left $2 \times 2$ block is the identity matrix because the
%  corresponding blocks of $A^T_n \R_n$ and of $A_{n-1}$ describe rotations
%  of opposite angles. We also note that the 33 entry of $\R^*_n$ is the
%  same of $\R_n$. Moreover, working out the off-diagonal terms 
%  we find that the matrix \eqref{app:endR*} is actually antisymmetric
%  and it can be cast in the form
Equation~\eqref{app:define_R*} gives
\begin{equation}
{\vec R}^*_n =
\begin{pmatrix}
1 && 0 && -a\rho^*_n \\
0 && 1 &&  a\tau^*_n  \\
 a\rho^*_n && -a\tau^*_n &&
1 - \frac{a^2}{2} [(\tau^*_n)^2 + (\rho^*_n)^2])\\
\end{pmatrix},
\label{app:endR*+}
\end{equation}
where
\begin{eqnarray}
\tau^*_n  &\equiv& 
  \frac{ s_{n+1} - s_{n}}{a\omega_0} \,\tau_n 
+ \frac{ c_{n+1} - c_{n}}{a\omega_0} \,\rho_n
\label{app:tau*}
\\
\rho^*_n &\equiv& 
  \frac{ s_{n+1} - s_{n}}{a\omega_0} \,\rho_n
- \frac{ c_{n+1} - c_{n}}{a\omega_0} \,\tau_n ,
\label{app:rho*}
\end{eqnarray}
with
\begin{eqnarray}
c_n \equiv \cos (na\omega_0) &\qquad&
s_n \equiv \sin (na\omega_0) .
\end{eqnarray}
In the limit $a \to 0$ one has $c_{n+1} - c_{n} \sim {\cal O}(a^2)$ and
$s_{n+1} - s_{n} \approx a {\omega}_0$, hence $\tau^*_n \approx \tau_n$
and $\rho^*_n \approx \rho_n$ as expected. The matrix \eqref{app:endR*+}
is formally identical to \eqref{app:approx_Rn} with the fields $\tau$
and $\rho$ replaced by $\tau^*$ and $\rho^*$. The bending persistence
length is then given by the analogous of Eq.~\eqref{app:lB}
\begin{equation}
\frac{1}{l_\text{B}}
= \frac{a}{\pi m} \int_{-\pi/2}^{\pi/2} \frac{\sin^2 my}{\sin^2 y}
\, \frac{ \left\langle |\widetilde{\tau^*}_{q} |^2 +
|\widetilde{\rho^*}_{q} |^2 \right\rangle}{N} \, dy.
\label{app:lB*}
\end{equation}
%  Performing a Fourier transform of \eqref{app:tau*} and \eqref{app:rho*}
%  and squaring we get
Using~\eqref{app:tau*} and \eqref{app:rho*} the Fourier transforms $\widetilde{\tau^*}_{q}$ and $\widetilde{\rho^*}_{q}$ 
can be expressed in terms of the original fields. The calculation
of the averages in \eqref{app:lB*} gives
\begin{eqnarray}
%  \frac{1}{N} 
%  \left\langle \left| \widetilde{\tau}^*_ q \right|^2\right\rangle =
%  \frac{1}{N} 
%  \left\langle \left| \widetilde{\rho}^*_ q \right|^2\right\rangle =
%  \Psi_{q+\Delta q} + \Psi_{q-\Delta q}
\left\langle |\widetilde{\tau^*}_{q} |^2 + |\widetilde{\rho^*}_{q}
|^2 \right\rangle &=& \frac{1 - \cos(a \omega_0)}{a^2 \omega_0^2}
\left\langle 
|\widetilde{\tau}_{q+\Delta q}|^2 +
|\widetilde{\tau}_{q-\Delta q}|^2 
\right.
\nonumber \\
&+&
\left.
|\widetilde{\rho}_{q+\Delta q}|^2 +
|\widetilde{\rho}_{q-\Delta q}|^2 
\right\rangle,
\label{app:ave*}
\end{eqnarray}
where $\Delta q \equiv N a \omega_0/2\pi$ is the momentum shift associated
with the double helix periodicity and originates from the Fourier transforms
of $c_n$ and $s_n$ in \eqref{app:tau*} and \eqref{app:rho*}.  Combining
\eqref{app:lB*} and \eqref{app:ave*} one obtains the expression of the
persistence length \eqref{DNA:lB} given in the main text.

%%%%%%%%%%%%%%%%%%%%%%%%%%%%%%%%%%%%%%%%%%%%%%%%%%%%%%%%%%%%%%%%%%%%  
\begin{figure}[t]
\includegraphics[width=0.48\textwidth]{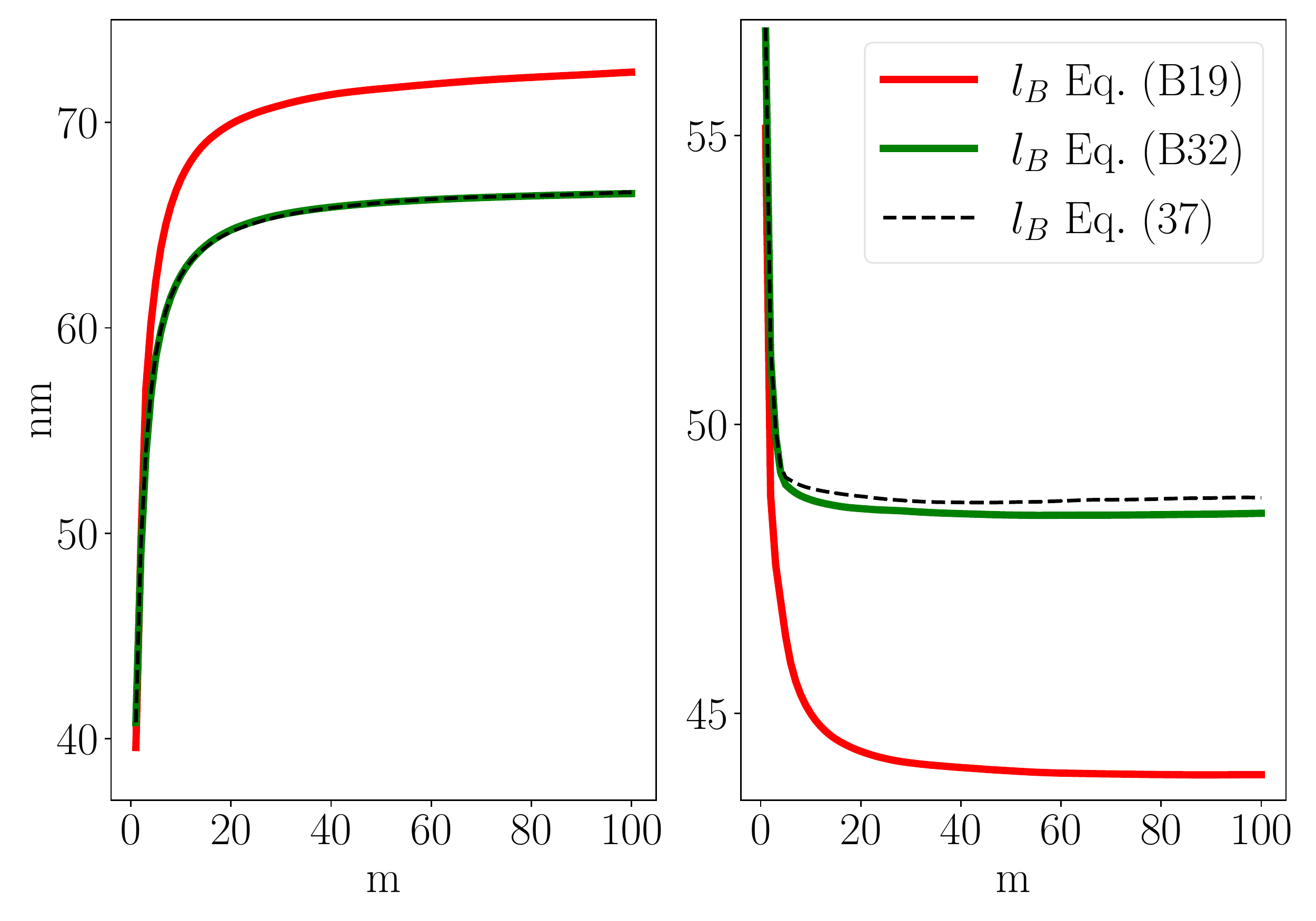}
\caption{Monte Carlo simulations with positive (left) and negative
(right) off-diagonal couplings. In both cases couplings between
step-parameters up to 2 steps displaced were included. The black
lines show the bending persistence length as deduced directly from the
tangent-tangent correlation function (Eq.~\eqref{DNA:deflB}). Indicated
in red is the expression derived for infinitesimal rotations
(Eq.~\eqref{app:lB}) and in green the improved expression
(Eq.~\eqref{app:ave*}).}
\label{fig:MC_lb} 
\end{figure}
%%%%%%%%%%%%%%%%%%%%%%%%%%%%%%%%%%%%%%%%%%%%%%%%%%%%%%%%%%%%%%%%%%%%

%%%%%%%%%%%%%%%%%%%%%%%%%%%%%%%%%%%%%%%%%%%%%%%%%%%%%%%%%%%%%%%%%
\begin{table}[b]
\caption{Parameters, given in nm, used in the Monte Carlo simulations 
for the calculation of $l_\text{B}$ shown in Fig.~\ref{fig:MC_lb}
($X_k$ indicates the coupling between site $n$ and $n+k$). For the
intrinsic twist density and discretization length $\omega_0 =
1.77$ nm$^{-1}$ and $a=0.34$ nm were used respectively.}
\begin{ruledtabular}
\begin{tabular}{l|ccc|ccc}
&& Simulation 1 &&& Simulation 2 & \\
& $X_0$ & $X_1$ & $X_2$ & $X_0$ & $X_1$ & $X_2$ \\
\hline
$A^t$ & 60 & 15 & 5 & 70 & -10 & -5 \\
$A^r$ & 40 & 8 & 4 & 60 & -10 & -4 \\
$C$   & 80 & 11 & 3 & 100 & -20 & -5 \\
$G$   & 20 & 2 & 1 & 30 & -10 & -5 \\
$A^{tr}$ & 0 & -2 & 0.5 & 0 & 0 & 0 \\
$B$ & 0 & 1 & 0.5 & 0 & 0 & 0 \\
\end{tabular}
\end{ruledtabular}
\label{app:tabel_MC}
\end{table}
%%%%%%%%%%%%%%%%%%%%%%%%%%%%%%%%%%%%%%%%%%%%%%%%%%%%%%%%%%%%%%%%%

In order to compare the quality of these approximations we
employed the Monte Carlo method used in \cite{nomi19} to generate
canonical ensembles of triads, distributed according to the free
energy~\eqref{dna:non_local}. In Figure~\eqref{fig:MC_lb} we compare
the direct calculation of the persistence length, as deduced from the
tangent-tangent correlation function (Eq.~\eqref{DNA:deflB}), with
the two approximations (Eq.~\eqref{app:lB} and Eq.~\eqref{app:ave*})
for two different set of model parameters (parameters given in Table
\ref{app:tabel_MC}). In both cases the expression that takes the
twist-dominance into account (Eq.~\eqref{app:ave*}), yields excellent
agreement with the direct calculation.

\section{Details all atom simulations}
\label{app:all_atom}

Using the x3dna webtool \cite{x3dna} we created an ideal B-DNA duplex
structure for various oligomers of 21 and 32 basepair length. All
sequences used in this work are listed in Table~\ref{tabel:aa_details}.
The structure was placed in a periodic dodecahedral box with at
least 1 nm distance between DNA and box boundary, followed by the
addition of water and 150 mM NaCl, resulting in a charge-neutral
system.  Preparation of the system consisted of energy minimization
(conjugate gradient with a force threshold of 100 kJ/mol nm) and a 100
ps position restrained molecular dynamics (MD) run, with restraints
on the DNA heavy atoms using a force constant of 1000 kJ/mol nm
in each direction. We used the parmbsc1 force field \cite{ivan16}
to describe the interactions between atoms, in combination with the
TIP3P water model \cite{jorg83}.  Non bonded interactions were treated
with a cut-off at 1.1 nm, and long range electrostatics were handled
by the Particle Mesh Ewald method. After equilibration, we performed
unrestrained molecular dynamics runs at constant temperature and pressure.
The velocity-rescaling thermostat \cite{buss07} kept the temperature
constant at 298 K and the Parrinello-Rahman barostat \cite{parr81}
kept the pressure constant at 1 bar. All molecular dynamics simulations
were performed with GROMACS version 2018.6 \cite{gromacs}. Frames were
stored every 1 ps. The rotational degrees of freedom of the inter-basepair
parameter - tilt, roll and twist - were then calculated with the Curves+
algorithm~\cite{lave09}.  Figure~\ref{fig:stiff_fourier_aa_all} shows
the elements of the stiffness matrix $\widetilde{M}_q$ for the $10$
different sequences with $N=15$ and the $3$ sequences with $N=27$
(corresponding to the $21$-mer and $32$-mer respectively), showing some
characteristic sample to sample variability.  The averages of these data
are shown in Fig.~\ref{fig:stiff_fourier-allatom}(a) and (b).

%  \esr{Should we not also include the $60$-mer simulation?}

%%%%%%%%%%%%%%%%%%%%%%%%%%%%%%%%%%%%%%%%%%%%%%%%%%
\begin{table}[t]
\caption{Details of the conducted simulations. N is 
the amount of deformation vectors $\vec{\Delta}_n$ 
considered per snapshot. }
\begin{ruledtabular}
\begin{tabular}{l | r | r}
sequence & simulation time (ns) & N\\
\hline
cgcattgcatacacttggacg & 1000 & 15 \\
cggtaccggctctggtcgccg & 1000 & 15 \\
cgcgatagcgttgtctcaccg & 1000 & 15 \\
cgagttttgaatataagctcg & 1000 & 15 \\
cgggatcaggaaggtggcccg & 1000 & 15 \\
cgttaaagaacatctacgtcg & 1000 & 15 \\
cgatgggcgcggaggcagccg & 1000 & 15 \\
cgtcgagtaacccctaattcg & 1000 & 15 \\
cggcacgggacgaaatcggcg & 1000 & 15 \\
cgactagcatgactgtgcgcg & 1000 & 15 \\
cgttatgtcattataagctcaatgcttatacg & 255 & 27 \\
cgacgtattaccgtacgattggcactatcacg & 254 & 27 \\
cgaagcactgccggggatctgacatccgcgcg & 174 & 27 \\
\end{tabular}
\end{ruledtabular}
\label{tabel:aa_details}
\end{table}
%%%%%%%%%%%%%%%%%%%%%%%%%%%%%%%%%%%%%%%%%%%%%%%%%%

%%%%%%%%%%%%%%%%%%%%%%%%%%%%%%%%%%%%%%%%%%%%%%%%%%%%%%%%%%%%%%%%%%%%  
\begin{figure}[t]
\includegraphics[width=0.48\textwidth]{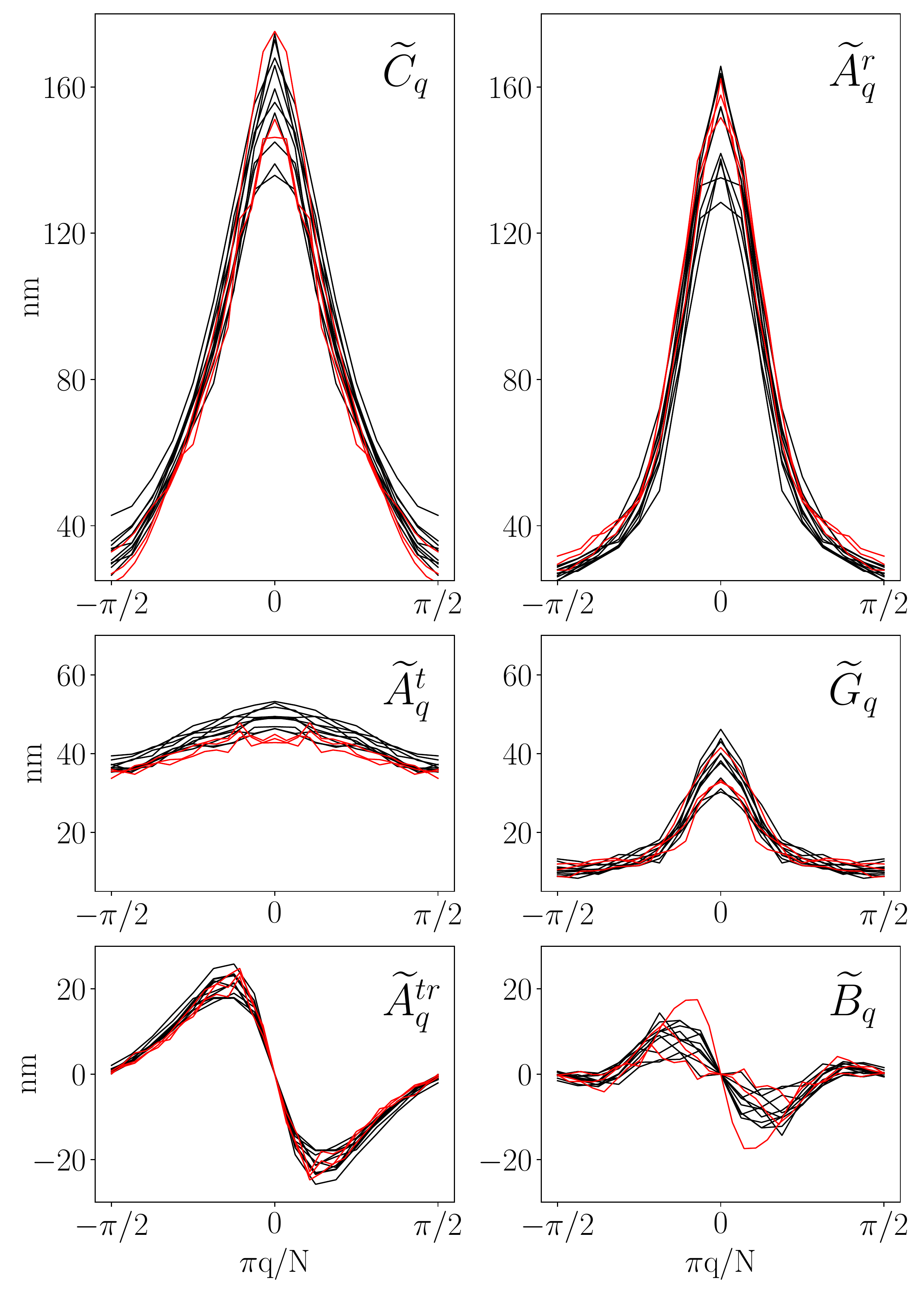}
\caption{Entries of the momentum space coupling matrices $\widetilde{M}_q$
for the full spectrum of rescaled momenta for all individual
simulations. Results of the $21$-mer ($N$$=$$15$) and $32$-mer
($N$$=$$27$) simulations are plotted in black and red respectively.}
\label{fig:stiff_fourier_aa_all} 
\end{figure}
%%%%%%%%%%%%%%%%%%%%%%%%%%%%%%%%%%%%%%%%%%%%%%%%%%%%%%%%%%%%%%%%%%%%

%  \bibliography{references}

\end{document}